\documentclass[11pt,a4paper]{article}
\usepackage[a4paper,margin=1.4in]{geometry}

\usepackage{amsmath}
\usepackage{amssymb}
\usepackage{amsfonts}
\usepackage{amsthm}
\usepackage{mathrsfs}

\usepackage{graphicx}
\usepackage{multirow}
\usepackage[title]{appendix}
\usepackage{xcolor}
\usepackage{textcomp}
\usepackage{manyfoot}
\usepackage{booktabs}

\usepackage{algorithm}
\usepackage{algorithmicx}
\usepackage{algpseudocode}

\usepackage{listings}

\usepackage{oubraces}
\usepackage{pifont}

\usepackage{tikz}
\usepackage{tikz-cd}
\usepackage{caption}
\usepackage{subcaption}

\usepackage[
    lambda,
    primitives,
    mm,
    ff,
    adversary,
    operators,
    advantage,
    events,
    asymptotics
]{cryptocode}

\usepackage{enumitem}

\usepackage[
    colorlinks=true,
    citecolor=blue,
    linkcolor=red,
    urlcolor=blue
]{hyperref}
\usepackage{cleveref}

\usepackage{authblk}

\colorlet{gc1}{blue!10}
\colorlet{gc2}{red!25}
\colorlet{gc3}{green!40!black!40}
\colorlet{gc4}{orange!90!black!40}

\newcommand{\A}{\mathcal{A}}
\newcommand{\B}{\mathcal{B}}
\newcommand{\C}{\mathcal{C}}
\newcommand{\D}{\mathcal{D}}

\renewcommand{\pp}{pp}

\newcommand{\sk}{\ensuremath{\mathsf{sk}}}
\newcommand{\pk}{\ensuremath{\mathsf{pk}}}

\renewcommand{\pp}{
    \text{\fontfamily{cmss}\selectfont pp}
}

\newcommand{\g}{\mathfrak{g}}
\renewcommand{\H}{\mathcal{H}}

\newcommand{\OSign}{\mathcal{O}_\mathsf{Sign}}
\newcommand{\OSimul}{\mathcal{O}_\mathsf{Simul}}
\newcommand{\OVerify}{\mathcal{O}_\mathsf{Verify}}

\newcommand{\Q}{\mathcal{Q}}

\newcommand{\Ocal}{\ensuremath{\mathcal{O}}}
\newcommand{\ZZ}{\ensuremath{\mathbb{Z}}}
\newcommand{\PP}{\ensuremath{\mathbb{P}}}

\newcommand{\Sign}{\sign}
\newcommand{\SKgen}{\ensuremath{\mathsf{SKeyGen}}}
\newcommand{\VKgen}{\ensuremath{\mathsf{VKeyGen}}}

\newcommand{\Exp}{\ensuremath{\mathsf{Exp}}}
\newcommand{\PSI}{\ensuremath{\mathsf{PSI}}}
\newcommand{\SDVS}{\ensuremath{\mathsf{SDVS}}}
\newcommand{\Simul}{\ensuremath{\mathsf{Simul}}}
\newcommand{\Verify}{\ensuremath{\mathsf{Verify}}}

\newcommand{\NT}{\ensuremath{\mathsf{NT}}}
\newcommand{\SUFCMA}{\ensuremath{\mathsf{SUF\text{-}CMA}}}
\newcommand{\ID}{\ensuremath{\mathsf{ID}}}

\renewcommand{\sample}{\ensuremath{\xleftarrow{\$}}}

\newcommand{\ppar}{\ensuremath{\mathsf{pp}}}
\newcommand{\oracles}{\ensuremath{\mathsf{oracles}}}

\newcommand{\cat}{~||~}
\newcommand{\GetRep}{\ensuremath{\mathsf{GetRep}}}

\newcommand{\Setup}{\ensuremath{\mathsf{Setup}}}
\newcommand{\Succ}{\ensuremath{\mathsf{Succ}}}
\newcommand{\Adv}{\ensuremath{\mathsf{Adv}}}
\newcommand{\st}{\ensuremath{\mathsf{st}}}

\theoremstyle{plain}

\newtheorem{theorem}{Theorem}

\newtheorem{lemma}[theorem]{Lemma}
\newtheorem{corollary}[theorem]{Corollary}

\theoremstyle{remark}

\newtheorem{remark}{Remark}
\newtheorem*{claim}{Claim}

\theoremstyle{definition}

\newtheorem{definition}{Definition}

\title{
    A Compact Post-quantum Strong Designated Verifier
    Signature Scheme from Isogenies
}

\author[1]{Farzin Renan}
\author[2]{Wendi Gao}
\author[2]{Jason T.~LeGrow}

\affil[1]{Independent Researcher}
\affil[2]{Department of Mathematics,
Virginia Polytechnic Institute and State University,
Blacksburg, Virginia, USA}

\affil[ ]{\texttt{farzin.renan@gmail.com},
\texttt{wendig@vt.edu},
\texttt{jlegrow@vt.edu}}

\date{}

\begin{document}

\maketitle

\begin{abstract}
\noindent Strong designated verifier signatures are a variant of digital signatures in which only one party---the designated verifier---is able to verify the authenticity of a signature. We propose a new strong designated verifier signature scheme based on (abelian) cryptographic group actions.

\vspace{1em}

\end{abstract}

\vspace{3em}

\noindent
\noindent\textbf{Acknowledgements.} Jason T.~LeGrow was partially supported by National Science Foundation grant NSF CNS 2528910. We would like to express our gratitude to Sarah~Arpin, Frederik Vercauteren, and the Isocrypt Study Group for their constructive feedback, which helped improve this paper. We are also especially grateful to P\'eter Kutas for his insightful discussions during the early stages of the draft.


\section{Introduction}

Digital signatures are cryptographic protocols used to ensure data integrity and authenticity. Ordinary digital signatures provide \emph{non-repudiation} by default: once Alice signs a message and publishes the signature, she can not later deny having signed the message, since anyone with access to her public key can verify that the signature is hers. While non-repudiation is a useful---or at worst, benign---feature in many applications, it is not always desirable. In 1996, Jakobsson, Sako, and Impagliazzo~\cite{DVS1} considered the task of authenticated off-the-record messaging, in which Alice wishes to authenticate herself and her messages to Bob, but in a way that does not allow Bob to convince any third party that the messages really came from Alice. Clearly, an ordinary digital signature would not work in this application. 

To achieve this, Jakobsson et al. \cite{DVS1} and Chaum \cite{DVS2} independently introduced the notion of \emph{Designated Verifier Signatures} (DVS) in 1996. In a DVS scheme, only a specific verifier---chosen by the signer at the time of signing---is capable of verifying the authenticity of a signature, while any third party remains unconvinced. This is achieved by ensuring that the designated verifier is capable of \emph{simulating} signatures that are computationally indistinguishable from authentic ones. When Bob receives a designated verifier signature from Alice, he knows that he did \emph{not} construct it, and so he is convinced it is valid. But if he sends the signature to a third party, that third party cannot tell whether it was produced by Alice (valid) or by Bob (invalid). This crucial feature---called \emph{non-transferability}---ensures that only the designated verifier is convinced of a signature's authenticity.

Building on preliminary work of Jakobson et al., in 2004 Laguillaumie and Vergnaud~\cite{SDVS} introduced \emph{Strong Designated Verifier Signatures} (SDVS) as an extension of designated verifier signatures that offers enhanced privacy guarantees. SDVS schemes provide additional privacy guarantees: in particular, they must satisfy the \emph{Privacy of Signer’s Identity} property, which says that given a valid signature and two candidate signing public keys, it is computationally infeasible for anyone other than the designated verifier to determine which key was used. 

\subsection{Related Work}

Strong designated verifier signature schemes have been extensively studied in both classical and post-quantum cryptographic settings.

\paragraph*{Classical Constructions.}  
Independent of the foundational work by Laguillaumie et al. \cite{SDVS}, Saeednia et al. \cite{SDVS2} proposed an efficient SDVS scheme by adapting the Schnorr signature and incorporating the designated verifier’s secret key into the verification process. Subsequently, Huang et al. \cite{shortSDVS} and Kang et al. \cite{idSDVS} introduced identity-based SDVS (IBSDVS) schemes and proved their unforgeability under the Bilinear Diffie–Hellman (BDH) assumption. Unfortunately, these constructions are rendered insecure in the quantum setting, due to Shor's algorithm \cite{Shor}.

\paragraph*{Post-Quantum Constructions.} There are a number of SDVS proposals from post-quantum assumptions.

Code-based SDVS constructions were proposed independently by Assar~\cite{assar2022}, Ren et al.~\cite{ren2022}, and Shooshtari et al. \cite{shooshtari2021}, based on the hardness of the bounded syndrome decoding problem. Unfortunately, it was shown in \cite{Thanalakshmi} that the constructions by Ren et al.~and Shooshtari et al.~fail to satisfy the critical non-transferability property, which is central to the SDVS notion. 

Wang et al. \cite{lattice1} propose a lattice-based SDVS using \emph{preimage sampleable functions} (PSFs) and Bonsai Trees. Their work builds on the framework of Gentry, Peikert and Vaikunthanathan~\cite{GPV}, which introduced PSFs as a primitive for lattice-based signatures. Although the construction offers provable security in the random oracle model, it suffers from efficiency issues due to large key and signature sizes. In 2016, Noh and Jeong \cite{lattice2} proposed a scheme in the standard model based on LWE-based encryption and a lattice-based chameleon hash function.
To improve practicality, Cai et al. \cite{lattice3} introduced a more efficient scheme based on the Ring-SIS assumption, using rejection sampling to reduce signature size while maintaining security. More recently, Zhang et al. \cite{lattice4} proposed an SDVS construction combining SIS and LWE assumptions, further contributing to the development of quantum-secure SDVS protocols. These lattice-based schemes incur key and signature sizes with complexity $\mathcal{O}(\lambda^2)$ \cite{lattice1,lattice2,lattice3,lattice4}, which may limit their suitability for resource-constrained environments.

In the isogeny-based setting, Sun et al.\cite{SIDH_SDVS} proposed an SDVS scheme built on the Supersingular Isogeny Diffie–Hellman (SIDH) protocol \cite{sidh}. This construction was rendered insecure by the landmark attacks against SIDH \cite{attack1SIDH, attack2SIDH, attack3SIDH} in 2022. In 2025, Khuc et al. \cite{khuc2025} proposed an SDVS scheme based on the CSIDH group action~\cite{csidh}, which is still believed to be secure.

\subsection{Our Contribution}

We present $\mathsf{CSI\text{-}SDVS}$, an efficient post-quantum SDVS scheme that also uses the CSIDH group action. Under the group action decisional Diffie-Hellman assumption (GA-DDH), the scheme satisfies all security properties required by an SDVS scheme, while offering very compact key and signature sizes and efficient signing and verification. We also prove that in the algebraic group action model (AGAM), the security of our scheme can be reduced to a computational (rather than decisional) assumption.

\paragraph*{Outline.} The next section recalls definition of an SDVS scheme and its security requirements. We then review the mathematical preliminaries and the group action-based hardness assumptions underlying our construction. In~\Cref{sec:CSI-SDVS}, we present our proposed scheme, $\mathsf{CSI\text{-}SDVS}$. We establish the security of our construction in the random oracle model (ROM) in~\Cref{sec:CSI-SDVS-security}, under the GA-DDH assumption. We then prove reductions to GAIP in Section~\ref{sec:AGAMProofs}. Finally, in~\Cref{sec:efficiency} we analyze the efficiency of the construction.

\section{Background}\label{sec:2-preliminaries}

\textbf{Notation.} We denote the security parameter by $\lambda$. For a finite set $\mathbb{X}$, the notation $x \xleftarrow{\$} \mathbb{X}$ indicates that $x$ is sampled uniformly at random from $\mathbb{X}$. A (deterministic or randomized) algorithm $A$ that outputs a value $x$ on input $y$ is denoted by $x \gets A(y)$. $\B^\A$ denotes that algorithm $\B$ has oracle access to algorithm $\A$ during its execution. A function $\negl[]: \mathbb{N} \rightarrow \mathbb{R}_{\ge 0}$ is said to be \emph{negligible} if for every positive polynomial $p(\cdot)$, there exists a threshold $n_0 \in \mathbb{N}$ such that for all $n \geq n_0$, it holds that $\negl[n] < \frac{1}{p(n)}$. The concatenation of two bit strings $s_1$ and $s_2$ is written as $s_1 \cat s_2$. For any positive integer $n$, we define $\mathbb{Z}_n := \mathbb{Z}/n\mathbb{Z}$.

\subsection{Strong Designated Verifier Signatures (SDVS)}\label{sec:SDVS}

An SDVS scheme involves two parties: a \emph{signer} $\mathsf{S}$ and a designated \emph{verifier} $\mathsf{V}$, who have secret key/public key pairs \( (\sk_S, \pk_S)\) and \((\sk_V, \pk_V)\), respectively. The signer $\mathsf{S}$ generates a signature $\sigma $ on a message $m$ using its secret key and the public key of the designated verifier. The resulting signature $\sigma$ can be verified only by $\mathsf{V}$ using its secret key $\sk_V$, ensuring that no third party can verify the signature or be convinced of its validity. The designated verifier is equipped with a simulation algorithm $\mathsf{Simul}$ that can generate a simulated signature $\sigma'$ on any message $m'$ using $\sk_V$ and the signer’s public key---the existence of the simulation algorithm prevents the designated verifier from convincing any third party of the validity of a signature, even if the verifier reveals their secret key.

The formal syntax of an SDVS scheme is presented below, following the framework introduced in~\cite{SIDH_SDVS}.

\begin{definition}[SDVS Scheme]\label{def:SDVS}
    A strong designated verifier signature scheme is defined by the following six polynomial-time algorithms:
    \begin{description}\setlength\itemsep{-.2em}
    \item[$\Setup(1^\lambda)\xrightarrow{\$}\mathsf{pp}$:] A $\mathsf{PPT}$ algorithm that, on inputs $1^\lambda$, outputs the public parameters $\mathsf{pp}$.
    \item[$\mathsf{SKeyGen}(\mathsf{pp})\xrightarrow{\$}(\sk_S, \pk_S)$:] A $\mathsf{PPT}$ algorithm that, on input $\mathsf{pp}$, generates a secret/public key pair $(\sk_S, \pk_S)$ for the signer $\mathsf{S}$.
    \item[$\mathsf{VKeyGen}(\mathsf{pp})\xrightarrow{\$}(\sk_V, \pk_V)$:] A $\mathsf{PPT}$ algorithm that, on input $\mathsf{pp}$, generates a secret/public key pair $(\sk_V, \pk_V)$ for the designated verifier $\mathsf{V}$.
    \item[$\Sign(\sk_S, \pk_V, m)\xrightarrow{\$}\sigma$:] A $\mathsf{PPT}$ (or $\mathsf{DPT}$) algorithm that, given the signer’s secret key $\sk_S$, the public key $\pk_V$ of the designated verifier, and a message $m \in \mathcal{M}$, outputs a real designated verifier signature $\sigma \in \mathsf{\Sigma}$.
    \item[$\Verify(\sk_V, \pk_S, m, \sigma)\rightarrow \{0,1\}$:] A $\mathsf{DPT}$ algorithm that, given the designated verifier’s secret key $\sk_V$, the signer’s public key $\pk_S$, a message $m \in \mathcal{M}$, and a signature $\sigma \in \mathsf{\Sigma}$, outputs a decision bit $b \in \{0,1\}$, where $b = 1$ indicates acceptance and $b = 0$ indicates rejection.
    \item[$\Simul(\sk_V, \pk_S, m)\xrightarrow{\$}\sigma'$:] A $\mathsf{PPT}$ (or $\mathsf{DPT}$) algorithm that, given the designated verifier’s secret key $\sk_V$, the signer’s public key $\pk_S$, and a message $m \in \mathcal{M}$, outputs a simulated designated verifier signature $\sigma' \in \mathsf{\Sigma}$.
\end{description}
\end{definition}

An SDVS scheme is required to satisfy the following notion of correctness. Let $\pp$ be the public parameters generated by $\Setup(1^{\lambda})$, and let $(\sk_S, \pk_S)$ and $(\sk_V, \pk_V)$ be the key pairs generated by $\mathsf{SKeyGen}(\pp)$ and $\mathsf{VKeyGen}(\pp)$, respectively. The correctness of an SDVS scheme requires that, for any message $m$, the following conditions hold with probability 1:
$$\Verify(\sk_V,  \pk_S, m, \Sign(\sk_S, \pk_V, m)) = 1,$$
for a real designated verifier signature, and
$$\Verify(\sk_V, \pk_S, m, \Simul(\sk_V, \pk_S, m)) = 1,$$
for a simulated designated verifier signature. These conditions ensure that honestly generated real and simulated designated verifier signatures are both valid and accepted by the verifier.

\subsection{Security Model}
An SDVS scheme must satisfy three security notions: Non-Transferability (NT), Strong Unforgeability under Chosen-Message Attacks (SUF-CMA), and Privacy of Signer's Identity (PSI). The formal security definitions corresponding to these properties are presented below.

\paragraph*{Non-Transferability (NT).}

In an SDVS scheme, the notion of non-transferability (NT) ensures that the designated verifier cannot convince any third party of the authenticity of a valid signature. This is achieved by enabling the verifier to use the $\Simul$ algorithm to generate simulated signatures that are computationally indistinguishable from genuine ones produced by the signer. As a result, any conviction about the validity of a signature cannot be reliably transferred to others. The NT security of our SDVS construction is formalized as follows.

\begin{definition}[NT Security]\label{def:NT-SDVS}
A strong designated verifier signature (SDVS) scheme is said to satisfy non-transferability (NT) if, for any $\mathsf{PPT}$ adversary $\A$, its advantage
$$
\Adv^{\NT}_{\SDVS,\A}(\lambda)
:=
\left|
\PP\!\left[\mathsf{Exp}^{\NT}_{\SDVS,\A}(\lambda)= 1\right]
- \frac{1}{2}
\right|,
$$
is negligible. Here, $\mathsf{Exp}^{\NT}_{\SDVS,\A}(\lambda)$ denotes the NT experiment between a challenger $\C$ and the adversary $\A$, depicted in Fig.~\ref{fig:NT}. The probability is taken over the randomness of $\A$ and the algorithms $\Sign$ and $\Simul$.

\begin{figure}[h!]
\centering
\begin{pchstack}[boxed, space=1em]
\procedure[linenumbering, lnstart=100]{$\Exp^{\NT}_{\SDVS,\adv}(1^\secpar)$}
{
    \ppar \gets \Setup(1^\secpar) \\
    (\sk_{S}, \pk_{S}) \gets \SKgen(\ppar)\\
    (\sk_{V}, \pk_{V}) \gets \VKgen(\ppar)\\
    m^* \gets \adv_1^{\Ocal_{\sign}, \Ocal_{\Simul}, \Ocal_{\Verify}}(\pk_S, \pk_V)\\
    \ID \sample \{S,V\} \\
    \pcif \ID = S \\
    \t \sigma^* \gets \sign(\sk_S,\pk_V,m^*) \\
    \pcelse \\
    \t \sigma^* \gets \Simul(\sk_V,\pk_S,m^*) \\
    \ID' \gets \adv_2^{\Ocal_{\sign}, \Ocal_{\Simul}, \Ocal_{\Verify}}(\pk_S, \pk_V, m^*, \sigma^*)\\
    \pcreturn [\ID = \ID']
}
\begin{pcvstack}[space=1em]
\procedure[linenumbering, lnstart=200]{$\Ocal_{\sign}(m)$}
{
    \sigma \gets \sign(\sk_S, \pk_{V}, m)\\
    \pcreturn \sigma
}
\procedure[linenumbering, lnstart=300]{$\Ocal_{\Simul}(m)$}
{
    \sigma \gets \Simul(\sk_V, \pk_{S}, m)\\
    \pcreturn \sigma
}
\procedure[linenumbering, lnstart=400]{$\Ocal_{\Verify}(m, \sigma)$}
{
    \pcreturn \Verify(\sk_V, \pk_S, m, \sigma)
}
\end{pcvstack}
\end{pchstack}
\caption{The Non-Transferability (NT) experiment.}
\label{fig:NT}
\end{figure}
\end{definition}

\paragraph*{Strong Unforgeability under Chosen-Message Attacks (SUF-CMA).}
This notion guarantees that no $\mathsf{PPT}$ adversary (without access to the secret key of either the signer or the designated verifier) can generate a valid designated verifier signature, even on messages that were previously queried via either the signing or simulating oracles. The formal definition is given below.

\begin{definition}[SUF-CMA Security]\label{def:SUF-SDVS}
A strong designated verifier signature (SDVS) scheme is said to be strongly unforgeable under chosen message attacks (SUF-CMA) if, for any $\mathsf{PPT}$ adversary $\A$, its advantage, defined by
\[
\Adv^{\mathsf{SUF\text{-}CMA}}_{\SDVS,\A}(\lambda)
=
\PP\!\left[
\mathsf{Exp}^{\mathsf{SUF\text{-}CMA}}_{\SDVS,\A}(\lambda) = 1
\right],
\]
is negligible. Here, $\mathsf{Exp}^{\mathsf{SUF\text{-}CMA}}_{\SDVS,\A}(\lambda)$ denotes the SUF-CMA experiment between a challenger~$\C$ and the adversary~$\A$, as illustrated in Fig.~\ref{fig:SUFCMA}.

\begin{figure}[h!]
\centering
\begin{pchstack}[boxed, space=2em]
\begin{pcvstack}[space=1em]
\procedure[linenumbering, lnstart=100]{$\Exp^{\SUFCMA}_{\SDVS,\adv}(1^\secpar)$}
{
    \ppar \gets \Setup(1^\secpar) \\
    (\sk_{S}, \pk_{S}) \gets \SKgen(\ppar)\\
    (\sk_{V}, \pk_{V}) \gets \VKgen(\ppar)\\
    (m^*,\sigma^*) \gets \adv^{\Ocal_{\sign}, \Ocal_{\Simul}, \Ocal_{\Verify}}(\pk_S, \pk_V)\\
    \pcassert (m^*,\sigma^*) \not\in \Q \\
    \pcreturn [\Verify(\sk_V, \pk_S, m^*, \sigma^*) = 1]
}

\procedure[linenumbering, lnstart=400]{$\Ocal_{\Verify}(m, \sigma)$}
{
    \pcreturn \Verify(\sk_V, \pk_S, m, \sigma)
}
\end{pcvstack}
\begin{pcvstack}[space=1em]
\procedure[linenumbering, lnstart=200]{$\Ocal_{\sign}(m)$}
{
    \sigma \gets \sign(\sk_S, \pk_{V}, m)\\
    \Q \gets \Q \cup \{(m,\sigma)\}\\
    \pcreturn \sigma
}

\procedure[linenumbering, lnstart=300]{$\Ocal_{\Simul}(m)$}
{
    \sigma \gets \Simul(\sk_V, \pk_{S}, m)\\
    \Q \gets \Q \cup \{(m,\sigma)\}\\
    \pcreturn \sigma
}
\end{pcvstack}
\end{pchstack}
\caption{The SUF-CMA experiment.}
\label{fig:SUFCMA}
\end{figure}

\noindent The probability is taken over the randomness of $\A$ and the algorithms $\Sign$ and $\Simul$.
\end{definition}

\paragraph*{Privacy of Signer's Identity (PSI).}

The privacy of signer's identity (PSI) security notion ensures that whenever there are two or more possible signers for a given message, the identity of the signer remains hidden. Within an SDVS scheme, even when a signature and two valid signing key pairs are available, no adversary observing the communication between the signer and the designated verifier can determine—except with negligible probability—which of the two secret keys was used to generate the signature. 

\begin{definition}[PSI Security]\label{def:PSI-SDVS}
A strong designated verifier signature (SDVS) scheme is said to satisfy privacy of the signer’s identity (PSI) if, for any $\mathsf{PPT}$ adversary $\A$, its advantage, defined by
$$
\Adv^{\mathsf{PSI}}_{\SDVS,\A}(\lambda)
:=
\left|
\PP\!\left[\mathsf{Exp}^{\mathsf{PSI}}_{\SDVS,\A}(\lambda)= 1\right]
- \frac{1}{2}
\right|,
$$
is negligible. Here, $\mathsf{Exp}^{\mathsf{PSI}}_{\SDVS,\A}(\lambda)$ denotes the PSI experiment between a challenger $\C$ and the adversary $\A$, as shown in Fig.~\ref{fig:PSI}. The probability is taken over the randomness of $\A$ and the algorithms $\Sign$ and $\Simul$.

\begin{figure}[h!]
\centering
\begin{pchstack}[boxed]
\procedure[linenumbering, lnstart=100]{$\Exp^{\PSI}_{\SDVS,\adv}(1^\secpar)$}
{
    \ppar \gets \Setup(1^\secpar) \\
    \pcfor i = 0,1\\
    \t (\sk_{S,i}, \pk_{S,i}) \gets \SKgen(\ppar)\\
    (\sk_{V}, \pk_{V}) \gets \VKgen(\ppar)\\
    b  \sample \{0,1\} \label{ln:fig_PSI_random_coin}\\
    m^* \gets \adv_1^{\Ocal_{\Simul}, \Ocal_{\Verify}}(\pk_{S,0}, \pk_{S,1}, \pk_V)\\
    \sigma^* \gets \sign(\sk_{S,b}, \pk_V, m^*)\\
    b' \gets \adv_2^{\Ocal_{\Simul}, \Ocal'_{\Verify}}(\pk_{S,0}, \pk_{S,1}, \pk_V, m^*, \sigma^*)\\
    \pcreturn [b = b']
}
\begin{pcvstack}[space=1em]
\procedure[linenumbering, lnstart=200]{$\Ocal_{\Simul}(i,m)$}
{
    \sigma \gets \Simul(\sk_V, \pk_{S,i}, m)\\
    \pcreturn \sigma
}
\procedure[linenumbering, lnstart=300]{$\Ocal_{\Verify}(i, (m, \sigma))$}
{
    \pcassert i \in \{0,1\}\\
    \pcreturn \Verify(\sk_V, \pk_{S,i}, m, \sigma)
}
\procedure[linenumbering, lnstart=400]{$\Ocal'_{\Verify}(i, (m, \sigma))$}
{
    \pcassert (m,\sigma) \neq (m^*, \sigma^*)\\
    \pcassert i \in \{0,1\}\\
    \pcreturn \Verify(\sk_V, \pk_{S,i}, m, \sigma)
}
\end{pcvstack}
\end{pchstack}
\caption{The Privacy of Signer's Identity (PSI) experiment.}
\label{fig:PSI}
\end{figure}
\end{definition}

\subsection{Cryptographic Group Actions}

We recall the notion of cryptographic group actions following~\cite[Section 3]{alamati_CGA}. Let \(\mathcal{G}\) be a group and \(\mathcal{X}\) be a set. A \emph{group action} is a triple \((\mathcal{G},\mathcal{X},\ast)\), where \(\ast : \mathcal{G} \times \mathcal{X} \to \mathcal{X}\) satisfies
\begin{enumerate}
    \item Identity: \(1_{\mathcal{G}} \ast X = X\) for all \(X \in \mathcal{X}\).
    \item Compatibility: \(g \ast (h \ast X) = (gh) \ast X\) for all \(g,h \in \mathcal{G}\) and all \(X \in \mathcal{X}\).
\end{enumerate}

The action is said to be \emph{free} if \(g \ast X \neq X\) for every \(g \in \mathcal{G} \setminus \{1_{\mathcal{G}}\}\) and every \(X \in \mathcal{X}\), and \emph{transitive} if for every \(X_1,X_2 \in \mathcal{X}\), there exists \(g \in \mathcal{G}\) such that \(g \ast X_1 = X_2\). Of particular importance in cryptographic applications are \emph{effective} group actions.

\begin{definition}[{Effective Group Action~\cite[Definition 3.4]{alamati_CGA}}]
A group action \( (\mathcal{G},\mathcal{X},\ast)\) is \emph{effective} if:
\begin{enumerate}
\item \(\mathcal{G}\) and \(\mathcal{X}\) are finite.
\item There are $\mathsf{PPT}$ algorithms for the following problems in \(\mathcal{G}\): membership testing, equality testing, sampling, the group operations, and inversion.
\item There is a $\mathsf{PPT}$ algorithm for membership testing in \(\mathcal{X}\), and each element of \(\mathcal{X}\) admits an efficiently-computable unique representaiton.
\item There is a distinguished element \(X_0\) of \(\mathcal{X}\)---the \emph{origin}---which is publicly-known.
\item There is a $\mathsf{PPT}$ algorithm for computing \(g \ast X\) given any \(g \in \mathcal{G}\) and \(X \in \mathcal{X}\).
\end{enumerate}
\end{definition}
The protocols in this work derive their security from the difficulty of three problems: the \emph{group action inverse problem} (GAIP), the \emph{group action computational Diffie-Hellman} (GA-CDH) problem, and the \emph{group action decisional Diffie-Hellman} (GA-DDH) problem. These problems are the group action analogues of the classical discrete logarithm problem, computational Diffie-Hellman problem, and decisional Diffie-Hellman problem, respectively.

\begin{definition}[Group Action Inverse Problem (GAIP)] 
\label{def:GAIP}
Let \( (\mathcal{G},\mathcal{X},\ast)\) be an effective group action with origin \(X_0\). Let \(g \sample \mathcal{G}\), and \(X_1 := g \ast X_0\). The GAIP problem is, given \( (X_0, X_1)\) to find \(g' \in \mathcal{G}\) such that \(g' \ast X_0 = X_1\).
\end{definition}

\begin{remark}
We note that when \( (\mathcal{G},\mathcal{X},\ast)\) is free, the solution to GAIP is unique.
\end{remark}

\begin{definition}[Group Action Computational Diffie-Hellman (GA-CDH) Problem]\label{def:CDH}
Let \( (\mathcal{G},\mathcal{X},\ast)\) be an effective group action with origin \(X_0\). Let \(g,h \sample \mathcal{G}\), \(X_1 := g \ast X_0\), and \(X_2 := h \ast X_0\). The GA-CDH problem is, given \( (X_1, X_2)\), to find \( (gh) \ast X_0\).
\end{definition}

\begin{definition}[Group Action Decisional Diffie-Hellman  (GA-DDH) Problem]\label{def:DDH} 
Let \( (\mathcal{G},\mathcal{X},\ast)\) be an effective group action with origin \(X_0\). Let \(g,h \sample \mathcal{G}\), \(X_1 := g \ast X_0\), and \(X_2 := h \ast X_0\). Let \(X_3\) be constructed as follows: sample \(b \in \{0,1\}\) and \(z \in \mathcal{G}\) uniformly at random, and define 

\[
X_3 = \begin{cases} 
    (gh) \ast X_0 & \mbox{if } b = 1 \\ 
    z \ast X_0 & \mbox{if } b = 0 
        \end{cases}
\] 
The GA-DDH problem is, given \( (X_1, X_2, X_3)\), to determine the value of \(b\).
\end{definition}

An effective group action is said to be a \emph{cyclic effective group action} (of known order) if there exists a group isomorphism $\iota \colon \ZZ_N \to \mathcal{G}$ for which both $\iota$ and $\iota^{-1}$ are efficiently computable. We recall the Algebraic Group Action Model (AGAM) for such group actions in~\Cref{sec:AGAM}; it forms the basis for the SUF-CMA and PSI security proofs presented in~\Cref{sec:AGAMProofs}.

\subsubsection{The Algebraic Group Action Model}
\label{sec:AGAM}

The algebraic group action model (AGAM) was introduced by Duman et
al.~\cite{AGAM} as a generalization of the algebraic group model (AGM)~\cite{AGM} to the group action setting. In this model, we
restrict our attention to \emph{algebraic} adversaries, as defined
in~\Cref{def:AlgebraicAdversary}.

\begin{definition}[Algebraic Adversary]
\label{def:AlgebraicAdversary}
Let $(\mathcal{G},\mathcal{X},\ast)$ be a cyclic effective group action, let $\Exp$ be a security experiment, and let $\A$ be an adversary for $\Exp$. Suppose that $\A$ has received the set elements $(X_i)_{i=0}^{n} \in \mathcal{X}^{n+1}$ so far. We say that $\A$ is \emph{algebraic} if, for every $Y \in \mathcal{X}$ appearing in an output of $\A$ or in an oracle query issued by $\A$, it additionally provides a pair $(a,i) \in \ZZ_N \times \{0,\ldots,n\}$ such that $Y = \iota(a) \ast X_i$.
\end{definition}

\noindent\textbf{Notation.} For a set element $Y$ output by $\A$ or used in an oracle query issued by $\A$, let
\[
(a,i) \gets \GetRep(Y)
\]
denote the representation provided by $\A$ for that occurrence of $Y$; that is, $\iota(a) \ast X_i = Y$, where $X_i$ is a set element previously received by $\A$ and $i \in \{0,\ldots,n\}$. If $Y$ is neither output by $\A$ nor used in an oracle query issued by $\A$, we define $\GetRep(Y) := \bot$.

\subsection{Elliptic Curves, Isogenies, and the CSIDH Group Action}

Here we recall some fundamental properties of elliptic curves. For a comprehensive treatment of the topic, see~\cite{washington,silverman}.

Let $k := \mathbb{F}_q$ denote a finite field, where $q := p^n$ for a prime $p > 3$ and a positive integer $n$. An \emph{elliptic curve} $E$ over $k$ is a smooth, projective, genus-1 curve defined over $k$ with a distinguished rational point $\mathcal{O}_E$ serving as the identity element for the group structure on $E$. For a positive integer $\ell$, the \emph{$\ell$-torsion subgroup} of $E$ is defined as $E[\ell] := \{P \in E(\overline{k}) \mid [\ell]P = \mathcal{O}_E\}$. An elliptic curve is called \emph{supersingular} if its group of $p$-torsion points is trivial over $\overline{\mathbb{F}}_p$, i.e., $E[p] = \{\mathcal{O}_E\}$. In particular, if $E/\mathbb{F}_p$ is supersingular, then $\#E(\mathbb{F}_p) = p + 1$.

An \emph{isogeny} between elliptic curves is a surjective morphism that preserves the group structure. Two curves $E_1$ and $E_2$ defined over $\mathbb{F}_q$ are said to be \emph{isogenous} if there exists an isogeny between them. For an isogeny $\phi: E \to E'$, its \textit{degree}, denoted $\deg(\phi)$, is the degree of the function field extension $[k(E) : \phi^*(k(E'))]$, where $\phi^{\ast}$ is the pullback of $\phi$ defined as $\phi^\ast: k(E')\rightarrow k(E)$, with $\phi^\ast(f) := f \circ \phi$ for $f \in k(E')$. We call $\phi$ \textit{separable} if the extension $k(E)/\phi^*(k(E'))$ is separable, a property ensuring that its degree exactly equals the size of its kernel, $\deg(\phi) = \#\ker(\phi)$. Consequently, for any finite subgroup $G \subset E(\mathbb{F}_q)$, there exists a unique separable isogeny (up to isomorphism) $\phi: E \to E' := E/G$ such that $\ker(\phi) = G$.

An \emph{endomorphism} is an isogeny from an elliptic curve $E$ to itself. 
The set of all endomorphisms of $E$, along with the zero map, forms a ring known as the \emph{endomorphism ring} of $E$, denoted $\mathrm{End}(E)$. If \(E\) is a supersingular elliptic curve defined over $\mathbb{F}_p$, the ring of $\mathbb{F}_p$-rational endomorphisms, denoted by $\mathrm{End}_{p}(E)$, is isomorphic to an order in an imaginary quadratic field. Following~\cite{csidh}, we choose parameters such that \(\mathrm{End}_p(E)\) is an order in \(\mathbb{Q}(\sqrt{-p})\).

Let $\mathfrak{O} \subset \mathbb{Q}(\sqrt{-p})$ be an order. The \emph{ideal class group} $\mathcal{C}\ell(\mathfrak{O})$ is defined as the quotient $\mathcal{I}_{\mathfrak{O}} / \mathcal{P}_{\mathfrak{O}}$, where $\mathcal{I}_{\mathfrak{O}}$ is the group of invertible fractional ideals of \(\mathfrak{O}\) and $\mathcal{P}_{\mathfrak{O}}$ is the subgroup of principal fractional ideals. The class group $\mathcal{C}\ell(\mathfrak{O})$ acts on the set $\mathcal{E}\ell\ell_{\mathbb{F}_{p}}(\mathfrak{O})$ of $\mathbb{F}_p$-isomorphism classes of supersingular elliptic curves defined  over $\mathbb{F}_p$ with endomorphism ring isomorphic to $\mathfrak{O}$ as follows. Let \([\mathfrak{a}] \in \mathcal{C}\ell(\mathfrak{O})\) be an ideal class, and suppose without loss of generality that \(\mathfrak{a}\) is integral. Let \(E\) be an elliptic curve with \(\mathrm{End}(E) \cong \mathfrak{O}\). Through this isomorphism we can view the elements of \(\mathfrak{a}\) as endomorphisms of \(E\), and hence define the subgroup
\[
S_{\mathfrak{a}} := \bigcap_{\alpha \in \mathfrak{a}} \ker(\alpha).
\]
This \(S_{\mathfrak{a}}\) is a finite subgroup of \(E\), and hence it is the kernel of a unique (up to \(\mathbb{F}_p\)-isomorphism) isogeny \(\psi_{\mathfrak{a}}\). We define the action \(
\ast : \mathcal{C}\ell(\mathfrak{O}) \times \mathcal{E}\ell\ell_{\mathbb{F}_{p}} (\mathfrak{O})\rightarrow \mathcal{E}\ell\ell_{\mathbb{F}_{p}}(\mathfrak{O})\) by
\[
[\mathfrak{a}] \ast E := \psi_\mathfrak{a}(E).
\]
This action is both free and transitive. Moreover, recent work~\cite{PEGASIS} demonstrates that this group action is effective\footnote{Earlier works using this group action do not require the group action to be effective; for many protocols, the weaker notion of \emph{restricted} effective group action~\cite[Definition 3.8]{alamati_CGA} is sufficient.}, and the relevant computational problems (GAIP, GA-CDH, and GA-DDH) are generally believed to be quantum-safe. We call this the Commutative Supersingular Isogeny Diffie-Hellman (CSIDH) group action, after the paper in which it first appeared~\cite{csidh} in the cryptographic context.

\section{Commutative Supersingular Isogeny Strong Designated Verifier Signatures: \textsf{CSI-SDVS}}\label{sec:CSI-SDVS}

We present our commutative supersingular isogeny-based strong designated verifier signature scheme, denoted by $\mathsf{CSI}\text{-}\mathsf{SDVS}$, built over a CSIDH-style group action. In Fig.~\ref{fig:OurProt}, we introduce the $\mathsf{CSI}\text{-}\mathsf{SDVS}$. Our scheme's public parameter setup follows that of CSIDH~\cite{csidh}, which we include here for completeness. A trusted party executes the setup procedure to generate the public parameters $\mathsf{pp}$ as follows. First, it selects a large prime $p = 4\ell_1 \ell_2 \dots \ell_n - 1$, where the $\ell_i$ are small, distinct, odd primes. CSIDH-512 takes $n = 74$, $\ell_1 , \hdots, \ell_{73}$ to be the first 73 odd primes, and $\ell_{74} = 587$. Next, it fixes a base curve $E: y^2 = x^3 + x$ defined over $\mathbb{F}_p$, such that $E \in \mathcal{E}\ell\ell_{\mathbb{F}_p}(\mathfrak{O})$, and chooses a generator $\mathfrak{g}$ of the class group $\mathcal{G} := \mathcal{C}\ell(\mathfrak{O})$ with class number $N\approx p^{1/2}$, where $\mathfrak{O} := \mathbb{Z}[\sqrt{-p}]$. It then samples a cryptographically secure hash function $\mathcal{H} : \{0,1\}^{\ast} \rightarrow \{0,1\}^{2\lambda}$, where $\lambda$ is a security parameter---in the security proof, we model this as a random oracle. Finally, it outputs the public parameters $\ppar := (p, \mathfrak{g}, N, E, \mathcal{H})$.

The remaining algorithms of our protocol are described in Fig.~\ref{fig:OurProt}.

\newcommand{\prot}{\mathsf{CSI\text{-}SDVS}}

\begin{figure}[h!]
\centering

\begin{pcvstack}[boxed, space=1em]

\begin{pchstack}[space=1em]
\procedure[lnstart=100,linenumbering]{$\prot.\SKgen(\ppar)$}
{
    \pcparse (p,[\g],N,E,\H) \gets \ppar \\
    \sk_S \sample \ZZ_N \\
    \pk_S \gets [\g^{\sk_S}] \ast E\\
    \pcreturn (\sk_S, \pk_S)
}
\procedure[lnstart=400,linenumbering]{$\prot.\Simul(\sk_V, \pk_S, m)$}
{
    z \sample \ZZ_N \\
    Y \gets [\g^{z+\sk_V}] \ast \pk_S \\
    h \gets \H(Y \,\|\, m)\\
    \pcreturn \sigma := (h,z)
}

\end{pchstack}

\begin{pchstack}[space=1em]
\procedure[lnstart=200,linenumbering]{$\prot.\VKgen(\ppar)$}
{
    \pcparse (p,[\g],N,E,\H) \gets \ppar \\
    \sk_V \sample \ZZ_N \\
    \pk_V \gets [\g^{\sk_V}] \ast E\\
    \pcreturn (\sk_V, \pk_V)
}

\procedure[lnstart=500,linenumbering]{$\prot.\Verify(\sk_V,\pk_S,m,\sigma)$}
{
    \pcparse (h,z) \gets \sigma \\
    \pcreturn [h = \H([\g^{z+\sk_V}] \ast \pk_S \,\|\, m)]
}

\end{pchstack}

\begin{pchstack}[space=1em]
\procedure[lnstart=300,linenumbering]{$\prot.\sign(\sk_S, \pk_V, m)$}
{
    \rlap{\ensuremath{z \sample \ZZ_N}}{\phantom{\ensuremath{\pcparse (p,[\g],N,E,\H) \gets \ppar}}} \\
    Y \gets [\g^{z+\sk_S}] \ast \pk_V \\
    h \gets \H(Y\,\|\,m)\\
    \pcreturn \sigma := (h,z)
}
\procedure[lnstart=600,linenumbering]{$\prot.\Verify'(\sk_S,\pk_V,m,\sigma)$}
{
    \pcparse (h,z) \gets \sigma \\
    \pcreturn [h = \H([\g^{z+\sk_S}] \ast \pk_V] \,\|\, m)]
}
\end{pchstack}

\end{pcvstack}
\caption{\(\prot\): Commutative Supersingular Isogeny Strong Designated Verifier Signatures.}
\label{fig:OurProt}
\end{figure}

The correctness of our scheme follows directly from the commutativity of the underlying class group action, since
\[
    [\mathfrak{g}^{z + \sk_V}] \ast \pk_S = [\mathfrak{g}^{z + \sk_S}] \ast \pk_V
\]
for any \(\sk_S, \sk_V \in \ZZ_N\).

Although the correctness definition only requires that the verifier be able to verify signatures from the signer, we note that the signer can verify by checking
$$
h \stackrel{?}{=} \mathcal{H}\big([\mathfrak{g}^{z + \sk_S}] \ast \pk_V \,\|\, m\big).
$$
We denote this ``signer-side verification'' as \(\Verify'\), and we include it in Fig.~\ref{fig:OurProt} to simplify notation in later proofs.

\section{Security Proofs} \label{sec:CSI-SDVS-security}

In this section, we establish the security properties of the proposed scheme, $\prot$. We first prove that that is satisfies NT security unconditionally, and we then establish its SUF-CMA and PSI security properties by reducing them to the GA-DDH assumption in the ROM. In the next section, we show that $\prot$ achieves SUF-CMA security and PSI under a computational (rather than decisional) assumption, in the algebraic group action model (AGAM).


\begin{theorem} 
\label{thm:Identical}
For the scheme $\prot$, for all public parameters $\pp$, valid signer and
designated verifier key pairs $(\sk_S,\pk_S)$ and $(\sk_V,\pk_V)$, and every
message $m \in \mathcal{M}$, the outputs of $\prot.\sign(\sk_S,\pk_V,m)$ and
$\prot.\Simul(\sk_V,\pk_S,m)$ are identically distributed.

\begin{proof}
Outputs of \(\prot.\sign(\sk_S, \pk_V, m)\) and \( \prot.\Simul(\sk_V, \pk_S, m)\) take the form \( (h,z)\), where \(h \in \{0,1\}^{2\lambda}\) and \(z \in \ZZ_N\). Note first that the distribution of the \(z\) component of both algorithms is identical: uniformly random in \(\ZZ_N\). For a fixed value of \(z\), the hash value computed in \(\prot.\sign(\sk_S, \pk_V, m)\) is
\[
h = \mathcal{H}\big([\g^{z + \sk_S}] \ast \pk_V \,\|\, m\big) = \mathcal{H}\big([\g^{z + \sk_S + \sk_V}] \ast E \,\|\, m\big)
\]
while in \(\prot.\Simul(\sk_V, \pk_S, m)\) it is
\[
h = \mathcal{H}\big([\g^{z + \sk_V}] \ast \pk_S \,\|\, m\big) = \mathcal{H}\big([\g^{z + \sk_V + \sk_S}] \ast E \,\|\, m\big).
\]
We see that the hash values are the same in both cases. Thus, the output distributions of the two algorithms (with a fixed message and fixed signer and verifier keys) are identical.
\end{proof}
\end{theorem}

\begin{corollary}
The SDVS scheme $\prot$ is NT secure.
\end{corollary}


\begin{theorem}
\label{thm:SUFCMA}
Under the GA-DDH assumption (Definition~\ref{def:DDH}), the proposed scheme, $\prot$, is SUF-CMA secure (as per Definition~\ref{def:SUF-SDVS}) in the
random oracle model.
\end{theorem}

The proof of~\Cref{thm:SUFCMA} follows by a standard sequence of games argument. At a high level, the games proceed as follows:

\begin{enumerate}[start=0, label = \ensuremath{\pcgame[\arabic*]}:, leftmargin=*, itemsep=0pt]
\item The SUF-CMA game, as illustrated in Fig.~\ref{fig:SUFCMA}.
\item The challenger chooses uniformly-random curve \(R\), and whenever an oracle would compute \( [\g^{z+\sk_V}] \ast \pk_S\) or  \( [\g^{z+\sk_S}] \ast \pk_V\) for some \(z \in \ZZ_N\), it instead computes \([\g^z] \ast R\). Under the GA-DDH assumption, any adversary's advantage changes only negligibly when moving from \(\pcgame[0]\) to \(\pcgame[1]\).
\item The game aborts and \(\A\) loses if two random oracle queries on different inputs return the same output. For any $\mathsf{PPT}$ adversary \(\A\), this happens with negligible probability by the birthday bound. 
\item The game aborts and \(\A\) loses if, when \(\A\) returns the forgery \( (m^*, \sigma^* = (h^*, z^*) )\), the random oracle has not been queried at \( ([\g^{z^*}] \ast R\,\|\,m^*)\). In \(\pcgame[2]\), such a forgery would pass verification only with negligible probability, so any adversary's advantage changes only negligibly when moving from \(\pcgame[2]\) to \(\pcgame[3]\).
\item The game no longer makes consistency checks between the oracle calls. We enumerate the ways that this could change the behaviour of the game, and show that for $\mathsf{PPT}$ adversaries, they all happen with negligible probability.
\end{enumerate}
Finally, in \(\pcgame[4]\), the adversary must make a query of the form \(\H([\g^{z^*}] \ast R\,\|\,m^*)\), where \( (m^*, \sigma^* = (h^*, z^*))\) is its forgery. In \(\pcgame[4]\), the adversary does not receive any information about \(R\)---either as input from the challenger or in the oracle responses---and so can make such a query with only negligible probability.

The first four games are depicted in Fig.~\ref{fig:SUFCMAG012}.

\begin{figure}[htbp!]
\centering

\begin{pcvstack}[boxed,space=1em]

\begin{pchstack}[space=1em]
\procedure[linenumbering, lnstart=100]{
    $\pcgame[0](1^\lambda)$
    \gamechange[gc1]{$\pcgame[1](1^\lambda)$}
    \gamechange[gc2]{$\pcgame[2](1^\lambda)$}
    \gamechange[gc3]{$\pcgame[3](1^\lambda)$}
}
{
    \Q \gets \varnothing, L \gets [~] \label{ln:SUFCMAinit}\\
    (\sk_S, \pk_S) \gets \prot.\SKgen(\ppar) \\
    (\sk_V, \pk_V) \gets \prot.\VKgen(\ppar) \\
    \gamechange[gc1]{$r \sample \ZZ_N$}\\
    \gamechange[gc1]{$R \gets [\g^r] \ast E$}\\
    \oracles \gets (\H,\OSign,\OSimul,\OVerify) \\
    (m^*, \sigma^* = (h^*,z^*)) \gets \A^{\oracles}(\pk_S,\pk_V) \\
    \gamechange[gc2]{$\pcassert \nexists x \neq x' \mbox{ s.t. } L[x] = L[x'] \neq \bot$}\label{ln:SUFCMAG2} \\
    \gamechange[gc3]{$\pcassert L[[\g^{z^*}] \ast R\,\|\,m^*] \neq \bot$} \label{ln:SUFCMAG3}\\
    \pcassert (m^*,\sigma^*) \not\in \Q  \\
    \gamechange[gc1]{$\pcif h^* = \H([\g^{z^*}] \ast R\,\|\,m^*)$} \label{ln:SUFCMAG0Ver}\\
    \gamechange[gc1]{$\t \pcreturn 1$}\\
    \gamechange[gc1]{$\pcreturn 0$}\\
    \pcreturn \prot.\Verify(\sk_V,\pk_S, m^*,\sigma^*)
}
\begin{pcvstack}[space=1em]

\procedure[lnstart=200,linenumbering]{$\OSign(m)$}
{
    z \sample \ZZ_N \\
    Y \gets [\g^{z+\sk_S}] \ast \pk_V \\
    \gamechange[gc1]{$Y \gets [\g^z] \ast R$} \\
    h \gets \H(Y\,\|\,m)\\
    \sigma \gets  (h,z) \\
    \Q \gets \Q \cup \{(m,\sigma)\} \\
    \pcreturn \sigma
}

\procedure[lnstart=300,linenumbering]{$\OSimul(m)$}
{
    z \sample \ZZ_N \\
    Y \gets [\g^{z+\sk_V}] \ast \pk_S \\
    \gamechange[gc1]{$Y \gets [\g^z] \ast R$} \\
    h \gets \H(Y\,\|\,m)\\
    \sigma \gets  (h,z) \\
    \Q \gets \Q \cup \{(m,\sigma)\} \\
    \pcreturn \sigma
}

\end{pcvstack}
\end{pchstack}
\begin{pchstack}[space=1em]
\procedure[linenumbering, lnstart=400]{$\H(x)$}
{
    \pcif L[x] = \bot \\
    \t L[x] \sample \{0,1\}^{2\secpar} \label{ln:SUFCMAHashDefine} \\
    \pcreturn L[x]
}
\procedure[lnstart=500,linenumbering]{$\OVerify(m,\sigma)$}
{
    \pcparse (h,z) \gets \sigma \\
    h' \gets \H([\g^{z+\sk_V}] \ast \pk_S\,\|\,m) \\
    \gamechange[gc1]{$h' \gets \H([\g^z] \ast R\,\|\,m)$} \\
    \pcreturn [h = h']
}

\end{pchstack}

\end{pcvstack}

\caption{Games \(\pcgame[0]\)--\(\pcgame[3]\) used in the proof of~\Cref{thm:SUFCMA}.}
\label{fig:SUFCMAG012}
\end{figure}

\begin{claim}
For any adversary \(\A\) running in time \(T\), there exists an algorithm \(\B\) running in time \(T + \poly[\lambda]\) such that 
\[
    \Adv^{\pcgame[0]}(\A) \leq \Adv^{\pcgame[1]}(\A) + \Adv^{\mathrm{GA\text{-}DDH}}(\B).
\]

\begin{proof}
Consider the GA-DDH adversary \(\B\) defined in Fig.~\ref{fig:SUFCMADDH}.

\begin{figure}[h!]
\centering

\begin{pchstack}[boxed,space=1em]
\begin{pcvstack}[space=1em]

\procedure[linenumbering, lnstart=100]{
    $\B^\A(X_1, X_2, X_3)$
}
{
    \Q \gets \varnothing, L \gets [~] \\
    (\pk_S, \pk_V, R) \gets (X_1, X_2, X_3) \\
    \oracles \gets (\H,\OSign,\OSimul,\OVerify) \\
    (m^*, \sigma^* = (h^*,z^*)) \gets \A^{\oracles}(\pk_S,\pk_V) \\
    \pcassert (m^*, \sigma^*) \not\in \Q\\
    \gamechange[white]{$\pcif h^* = \H([\g^{z^*}] \ast R\,\|\,m^*)$}\\
    \gamechange[white]{$\t \pcreturn 1$}\\
    \gamechange[white]{$\pcreturn 0$}
}

\begin{pchstack}[space=1em]

\procedure[linenumbering, lnstart=400]{$\H(x)$}
{
    \pcif L[x] = \bot \\
    \t L[x] \sample \{0,1\}^{2\secpar} \label{ln:SUFCMAHashDef} \\
    \pcreturn L[x]
}

\procedure[lnstart=500,linenumbering]{$\OVerify(m,\sigma)$}
{
    \pcparse (h,z) \gets \sigma \\
    h' \gets \H([\g^z] \ast R\,\|\,m) \label{ln:SUFCMAVerHash} \\
    \pcreturn [h = h']
}

\end{pchstack}
\end{pcvstack}
\begin{pcvstack}[space=1em]

\procedure[lnstart=200,linenumbering]{$\OSign(m)$}
{
    z \sample \ZZ_N \\
    Y \gets [\g^z] \ast R \\
    h \gets \H(Y\,\|\,m)\\
    \sigma \gets  (h,z) \\
    \Q \gets \Q \cup \{(m,\sigma)\} \\
    \pcreturn \sigma
}
\procedure[lnstart=300,linenumbering]{$\OSimul(m)$}
{
    z \sample \ZZ_N \\
    Y \gets [\g^z] \ast R \\
    h \gets \H(Y\,\|\,m)\\
    \sigma \gets  (h,z) \\
    \Q \gets \Q \cup \{(m,\sigma)\} \\
    \pcreturn \sigma
}

\end{pcvstack}

\end{pchstack}

\caption{A GA-DDH distinguisher \(\B\). Its input \( (X_1, X_2, X_3)\) is drawn uniformly at random from either the group action Diffie-Hellman distribution on \( (\mathcal{G},\mathcal{X},\ast)\), or the uniform distribution on \(\mathcal{X}^3\), each with probability \(\frac{1}{2}\).}
\label{fig:SUFCMADDH}
\end{figure}

Observe that when \( (X_1, X_2, X_3)\) is sampled from the Diffie-Hellman distribution on \( (\mathcal{G},\mathcal{X},\ast)\), \(\B\) is honestly playing \(\pcgame[0]\) with \(\A\), and when \( (X_1, X_2, X_3)\) is sampled uniformly from \(\mathcal{X}^3\), \(\B\) is honestly playing \(\pcgame[1]\) with \(\A\). Thus,
\[
    |\Adv^{\pcgame[0]}(\A) - \Adv^{\pcgame[1]}(\A)| \leq \Adv^{\mathrm{GA\text{-}DDH}}(\B).\qedhere
\]

\end{proof}
\end{claim}

\begin{claim}
For any algorithm \(\A\) that makes at most \(q\) total queries to \(\H, \OSign, \OSimul\), and \(\OVerify\), we have
\[
    \Adv^{\pcgame[1]}(\A) \leq \Adv^{\pcgame[2]}(\A) + \frac{q^2}{2^{2\lambda}}.
\]

\begin{proof}
The only difference between \(\pcgame[1]\) and \(\pcgame[2]\) is on line~\ref{ln:SUFCMAG2}, and so the difference in advantages is at most the probability that the assertion there fails. The assertion fails if and only if there have been distinct queries \(x\neq x'\) to \(\H\) (either explicit by the adversary, or implicit through queries to other oracles) such that \(\H(x) = \H(x')\). Since \(\H\) is queried at at most \(q\) inputs, and each output of \(\H\) is chosen independently and uniformly at random, this probability is bounded above by \(\frac{q^2}{2^{2\lambda}}\) by the birthday bound. The claim follows.
\end{proof}

\begin{claim}
For any algorithm \(\A\), we have
\(
    \Adv^{\pcgame[2]}(\A) \leq \Adv^{\pcgame[3]}(\A) + \frac{1}{2^{2\lambda}}.
\)

\begin{proof}
The only difference between \(\pcgame[2]\) and \(\pcgame[3]\) occurs on line~\ref{ln:SUFCMAG3}. The differences in advantages between the games is bounded by the probability that \(\A\) wins the game conditioned on the assertion failing; that is,
\[
    |\Adv^{\pcgame[2]}(\A) - \Adv^{\pcgame[3]}(\A)| \leq \PP[h^* = \H([\g^{z^*}] \ast R\,\|\,m^*) \mid L[ [g^{z^*}] \ast R\,\|\,m^*] = \bot].
\]
Since \(L[ [g^{z^*}] \ast R\,\|\,m^*] = \bot\), we know that \(\H\) has not been queried (directly or by a query to \(\OSign\), \(\OSimul\), or \(\OVerify\)) at \( ([g^{z^*}] \ast R\,\|\,m^*)\). Thus, when the challenger verifies \( (m^*, \sigma^*)\) on line~\ref{ln:SUFCMAG0Ver}, the value of \(\H(([g^{z^*}] \ast R\,\|\,m^*)\) will be chosen uniformly at random on line~\ref{ln:SUFCMAHashDefine}. Then, we have \(h^* = \H([\g^{z^*}] \ast R\,\|\,m^*)\) with probability \(\frac{1}{2^{2\lambda}}\), as required.
\end{proof}
\end{claim}

Finally, before proceeding to the proof of~\Cref{thm:SUFCMA}, we consider the $\pcgame[4]$, depicted in Fig.~\ref{fig:SUFCMAG4}.

\begin{figure}[htbp!]
\centering

\begin{pcvstack}[boxed,space=1em]
\begin{pchstack}
\procedure[linenumbering, lnstart=100]{
    $\pcgame[4](1^\lambda)$
}
{
    \Q \gets \varnothing, L \gets [~] \\
    (\sk_S, \pk_S) \gets \prot.\SKgen(\ppar) \\
    (\sk_V, \pk_V) \gets \prot.\VKgen(\ppar) \\
    \gamechange[white]{$r \sample \ZZ_N$}\\
    \gamechange[white]{$R \gets [\g^r] \ast E$}\\
    \oracles \gets (\H,\OSign,\OSimul,\OVerify) \\
    (m^*, \sigma^* = (h^*,z^*)) \gets \A^{\oracles}(\pk_S,\pk_V) \\
    \gamechange[white]{$\pcassert L[[\g^{z^*}] \ast R\,\|\,m^*] \neq \bot$} \\
    \gamechange[white]{$\pcassert \nexists x \neq x' \mbox{ s.t. } L[x] = L[x'] \neq \bot$} \\
    \pcassert (m^*,\sigma^*) \not\in \Q  \\
    \gamechange[white]{$\pcif h^* = \H([\g^{z^*}] \ast R\,\|\,m^*)$} \\
    \gamechange[white]{$\t \pcreturn 1$}\\
    \gamechange[white]{$\pcreturn 0$}
}
\begin{pcvstack}[space=1em]

\procedure[lnstart=200,linenumbering]{$\OSign(m),~\OSimul(m)$}
{
    z \sample \ZZ_N \\
    h \sample \{0,1\}^{2\lambda}\\
    \sigma \gets  (h,z) \\
    \Q \gets \Q \cup \{(m,\sigma)\} \\
    \pcreturn \sigma
}
\procedure[linenumbering, lnstart=300]{$\H(x)$}
{
    \pcif L[x] = \bot \\
    \t L[x] \sample \{0,1\}^{2\secpar} \\
    \pcreturn L[x]
}

\procedure[lnstart=400,linenumbering]{$\OVerify(m,\sigma)$}
{
    \pcreturn [(m,\sigma) \in \Q]
}
\end{pcvstack}
\end{pchstack}
\end{pcvstack}
\caption{The \(\pcgame[4]\) used in the proof of~\Cref{thm:SUFCMA}.}
\label{fig:SUFCMAG4}
\end{figure}
\end{claim}

\begin{claim}
For any algorithm \(\A\) that makes at most \(q_\H\) (respectively, \(q_\Sign, q_\Simul, q_\Verify\)) queries to \(\H\) (respectively \(\OSign, \OSimul, \OVerify\)), we have 
\begin{align*}
\Adv^{\pcgame[3]}(\A) \leq \Adv^{\pcgame[4]}&(\A) + \frac{(q_\Sign+q_\Simul+2q_\H)(q_\Sign+q_\Simul) + q_\H q_\Verify}{N}\\
&+ \frac{q_\H q_\Verify}{2^{2\lambda}} + \frac{q_\Verify(q_\Sign+q_\Simul)}{2^{2\lambda} N}.\\
\end{align*}

\begin{proof}
From the adversary's perspective, \(\pcgame[3]\) is identical to \(\pcgame[4]\) unless the result of an oracle query is contradicted by a later oracle query. In particular, the games are identical unless one of the following events happens in \(\pcgame[4]\):
\begin{description}[labelindent=0pt]
\item[\(\OSign/\OSimul\) contradicts \(\OSign/\OSimul\):] Distinct queries to either \(\OSign\) or \(\OSimul\) with the same message \(m\) have the same value of \(z\) and different values of \(h\).

Since each call to \(\OSign\) and \(\OSimul\) generates an independent, uniformly random value of \(z\), and there are at most \( q_\Sign + q_\Simul\) total queries to these oracles, this event happens with probability at most \(\frac{(q_\Sign+q_\Simul)^2}{N}\) by the birthday bound.

\item[\(\OVerify\) contradicts \(\OSign/\OSimul\):] \(\OVerify\) rejects a signature generated by \(\OSign\) or \(\OSimul\).

By construction, this never happens.

\item[\(\H\) contradicts \(\OSign/\OSimul\):] The adversary queries \(\OSign\) or \(\OSimul\) at \(m\) and it returns \( (h,z)\). Later, the adversary queries \(\H\) at \( ([\g^z] \ast R \,\|\, m)\) and it returns \(h' \neq h\).

Up until such a query to \(\H\) occurs, \(R\) is information-theoretically hidden from the adversary, and so such a query cannot be made except by random chance. There are \(N\) possible values of \(R\), and the adversary makes at most \(q_\H\) queries to \(\H\) and at most \( q_\Sign+q_\Simul\) queries to \(\OSign\) and \(\OSimul\) in total. Thus, this event occurs with probability at most \( \frac{q_H(q_\Sign+q_\Simul)}{N}\).

\item[\(\OSign/\OSimul\) contradicts \(\OVerify\):] The adversary queries \(\OVerify\) at some \( (m,\sigma = (h,z))\) and it is rejected; later, the adversary queries \(\OSign\) or \(\OSimul\) at \(m\) and the oracle returns \( (h,z)\).

Since each call to \(\OSign\) and \(\OSimul\) generates independent, uniformly random values of \(z\) and \(h\), and there are at most \( q_\Sign + q_\Simul\) total queries to \(\OSign\) and \(\OSimul\) and \(q_\Verify\) queries to \(\OVerify\), this event happens with probability at most \(\frac{q_\Verify(q_\Sign+q_\Simul)}{2^{2\lambda} N}\).

\item[\(\OVerify\) contradicts \(\OVerify\):] The adversary queries \(\OVerify\) at some \( (m, \sigma)\) twice and it is accepted once and rejected once.

This can only happen in the following way: \( (m, \sigma)\) is rejected because it is not in \(\Q\), then a query to \(\OSign\) or \(\OSimul\) adds \( (m, \sigma)\) to \(\Q\), and then the adversary queries \(\OVerify\) at \( (m, \sigma)\) again. This case is subsumed by the previous one.

\item[\(\H\) contradicts \(\OVerify\):] The adversary queries \(\OVerify\) at a pair \( (m, \sigma = (h,z)) \not\in \Q\) (and so, it is rejected). Later, the adversary queries \(\H([\g^z] \ast R\,\|\,m)\) and it returns \(h\).

Since \(\H([\g^z] \ast R\,\|\,m)\) is set uniformly at random on line~\ref{ln:SUFCMAHashDef}, and the adversary makes at most \(q_\H\) queries to \(\H\) and \(q_\Verify\) queries to \(\OVerify\), this occurs with probability at most \(\frac{q_\H q_\Verify}{2^{2\lambda}}\).

\item[\(\OSign/\OSimul\) contradicts \(\H\):] The adversary queries \(\H([\g^z] \ast R\,\|\,m)\) and it returns \(h\). Later, the adversary queries \(\OSign\) or \(\OSimul\) at \(m\) and the oracle returns \( \sigma = (h',z)\) with \(h' \neq h\).

Since each call to \(\OSign\) and \(\OSimul\) generates an independent, uniformly random value of \(z\), and there are at most \(q_\H\) queries to \(\H\) and \( q_\Sign + q_\Simul\) total queries to \(\OSign\) and \(\OSimul\), this event happens with probability at most \(\frac{q_\H(q_\Sign+q_\Simul)}{N}\).

\item[\(\OVerify\) contradicts \(\H\):] The adversary queries \(\H\) at some \( ([\g^z] \ast R\,\|\,m)\) and it returns \(h\). Later, the adversary queries \(\OVerify\) at \( (m, \sigma = (h,z))\) and the pair is rejected.

Up until such a query occurs, \(R\) is information-theoretically hidden from the adversary, and so such a query cannot be made except by random chance. There are \(N\) possible values of \(R\), and the adversary makes at most \(q_\H\) queries to \(\H\) and at most \( q_\Verify\) queries to \(\OVerify\). Thus, this event occurs with probability at most \( \frac{q_H q_\Verify}{N}\).

\item[\(\H\) contradicts \(\H\):] The adversary queries \(\H\) twice at the same input \(x\) and receives two different outputs. By construction, this never happens.
\end{description}

The result follows by summing the probabilities of these events. \qedhere
\end{proof}
\end{claim}

\begin{claim}
For any adversary \(\A\) that makes at most \(q_\H\) queries to \(\H\), we have \(\Adv^{\pcgame[4]}(\A) \leq \frac{q_\H}{N}\).

\begin{proof}
In \(\pcgame[4]\), each oracle's output is independent of the value of \(R\). Thus, for any \(z^*\), the probability that \(\H\) has been queried at \( ([\g^{z^*}] \ast R\,\|\,m)\) for some \(m\) is at most \(\frac{q_H}{N}\). If no such query has been made, then \(\A\) loses the game because of the assertion on line~\ref{ln:SUFCMAG3}. The result follows.
\end{proof}
\end{claim}

\begin{proof}[Proof (of~\Cref{thm:SUFCMA})]
Altogether, for any adversary \(\A\) running in time \(T\) and making at most \(q_\H\) (respectively, \(q_\Sign, q_\Simul, q_\Verify\)) queries to \(\H\) (respectively \(\OSign, \OSimul, \OVerify\)), there is an adversary \(\B\) that runs in time \(T + \poly[\lambda]\) such that
\begin{align*}
\Adv^{\SUFCMA}_{\prot}(\A) \leq \Adv^{\mathrm{DDH}}(\B) &+ \frac{q^2 + q_\H q_\Verify + 1}{2^{2\lambda}} + \frac{q_\Verify(q_\Sign+q_\Simul)}{2^{2\lambda} N} \\
&+ \frac{(q_\Sign+q_\Simul+2q_\H)(q_\Sign+q_\Simul) + q_\H q_\Verify}{N},
\end{align*}
where \(q = q_\H + q_\Sign + q_\Simul + q_\Verify\). Thus, when \(\A\) is $\mathsf{PPT}$ (so that \(T = \poly[\lambda]\) and \(q = \poly[\lambda]\)), \(\Adv^{\SUFCMA}_{\prot}(\A) = \negl[\lambda]\) under the GA-DDH assumption, as required.
\end{proof}

\begin{theorem}
\label{thm:PSI}
Under the GA-DDH assumption (Definition~\ref{def:DDH}), the proposed scheme, $\prot$, is PSI secure (as per Definition~\ref{def:PSI-SDVS}) in the
random oracle model.
\end{theorem}
Like the proof of~\Cref{thm:SUFCMA}, for~\Cref{thm:PSI} we proceed by a sequence of games.
\begin{enumerate}[start=0, label = \ensuremath{\pcgame[\arabic*]}:, leftmargin=*, itemsep=0pt]
\item The PSI~game, as shown in Fig.~\ref{fig:PSI}.
\item The challenger chooses uniformly-random curve \(R_0\), and whenever an oracle would compute \( [\g^{z+\sk_V}] \ast \pk_{S,0}\) or  \( [\g^{z+\sk_{S,0}}] \ast \pk_V\) for some \(z \in \ZZ_N\), it instead computes \([\g^z] \ast R_0\). Under the GA-DDH assumption, any adversary's advantage changes only negligibly when moving from \(\pcgame[0]\) to \(\pcgame[1]\).
\item The challenger chooses uniformly-random curve \(R_1\), and whenever an oracle would compute \( [\g^{z+\sk_V}] \ast \pk_{S,1}\) or  \( [\g^{z+\sk_{S,1}}] \ast \pk_V\) for some \(z \in \ZZ_N\), it instead computes \([\g^z] \ast R_1\). Under the GA-DDH assumption, any adversary's advantage changes only negligibly when moving from \(\pcgame[1]\) to \(\pcgame[2]\).
\item The game aborts and \(\A\) loses if two random oracle queries on different inputs return the same output. For any $\mathsf{PPT}$ adversary \(\A\), this happens with negligible probability by the birthday bound. 
\item The game no longer makes consistency checks between the oracle calls. We enumerate the ways that this could change the behaviour of the game, and show that for $\mathsf{PPT}$ adversaries, they all happen with negligible probability.
\end{enumerate}
Finally, we show that no adversary can win \(\pcgame[4]\) with probability different from \(\frac{1}{2}\), completing the proof.

Games \(\pcgame[0], \pcgame[1], \pcgame[2]\), and \(\pcgame[3]\) are depicted in Fig.~\ref{fig:PSIG0123}. These game hopes closely resemble ones made in the proof of~\Cref{thm:SUFCMA}, and so we omit many of the details.

\begin{figure}[ht!]
\centering
\begin{pchstack}[boxed, space=1em]
\procedure[linenumbering, lnstart=100]
{
    $\pcgame[0](1^\lambda)$
    \gamechange[gc1]{$\pcgame[1](1^\lambda)$}
    \gamechange[gc2]{$\pcgame[2](1^\lambda)$}
    \gamechange[gc3]{$\pcgame[3](1^\lambda)$}
}
{
    \ppar \gets \Setup(1^\secpar) \\
    \pcfor i = 0,1\\
    \t (\sk_{S,i}, \pk_{S,i}) \gets \SKgen(\ppar)\\
    (\sk_{V}, \pk_{V}) \gets \VKgen(\ppar)\\
    b  \sample \{0,1\} \\
    R_0 \gets [\g^{\sk_V}] \ast \pk_{S,0} \\
    R_1 \gets [\g^{\sk_V}]\ast \pk_{S,1} \\
    \gamechange[gc1]{$r_0 \sample \ZZ_N$} \\
    \gamechange[gc1]{$R_0 \gets [\g^{r_0}] \ast E$}\\
    \gamechange[gc2]{$r_1 \sample \ZZ_N$} \\
    \gamechange[gc2]{$R_1 \gets [\g^{r_1}] \ast E$}\\
    m^* \gets \adv_1^{\H,\Ocal_{\Simul}, \Ocal_{\Verify}}(\pk_{S,0}, \pk_{S,1}, \pk_V)\\
    z^* \sample \ZZ_N \\
    h^* \gets \H([\g^{z^*}] \ast R_b \,\|\,m^*) \\
    \sigma^* \gets (h^*, z^*)\\
    \gamechange[gc3]{$\pcassert \nexists x \neq x' \mbox{ s.t. } L[x] = L[x'] \neq \bot$}\label{ln:PSIG3} \\
    b' \gets \adv_2^{\H,\Ocal_{\Simul}, \Ocal'_{\Verify}}(\pk_{S,0}, \pk_{S,1}, \pk_V, m^*, \sigma^*)\\
    \pcreturn [b = b']
}
\begin{pcvstack}[space=1em]
\procedure[lnstart=200,linenumbering]{$\OSimul(i,m)$}
{
    z \sample \ZZ_N \\
    h \gets \H([\g^{z}] \ast R_i \,\|\,m)\\
    \sigma \gets  (h,z) \\
    \pcreturn \sigma
}
\procedure[linenumbering, lnstart=300]{$\H(x)$}
{
    \pcif L[x] = \bot \\
    \t L[x] \sample \{0,1\}^{2\secpar} \\
    \pcreturn L[x]
}
\procedure[lnstart=400,linenumbering]{$\OVerify(i, (m,\sigma))$}
{
    \pcassert i \in \{0,1\} \\
    \pcparse (h,z) \gets \sigma \\
    h' \gets \H([\g^z] \ast R_i \,\|\,m) \label{ln:PSIVerHash} \\
    \pcreturn [h = h']
}
\procedure[lnstart=500,linenumbering]{$\OVerify'(i, (m,\sigma))$}
{
    \pcassert (m,\sigma) \neq (m^*, \sigma^*) \\
    \pcassert i \in \{0,1\} \\
    \pcparse (h,z) \gets \sigma \\
    h' \gets \H([\g^z] \ast R_i \,\|\,m) \label{ln:PSIVerHash'} \\
    \pcreturn [h = h']
}
\end{pcvstack}
\end{pchstack}
\caption{The games \(\pcgame[0], \pcgame[1], \pcgame[2],\) and \(\pcgame[3]\) used in the proof of~\Cref{thm:PSI}.}
\label{fig:PSIG0123}
\end{figure}

\begin{claim}
For any adversary \(\A\) running in time \(T\), there exists an algorithm \(\B\) running in time \(T + \poly[\lambda]\) such that 
\[
    \Adv^{\pcgame[0]}(\A) \leq \Adv^{\pcgame[2]}(\A) + 2\Adv^{\mathrm{GA\text{-}DDH}}(\B).
\]

\begin{proof}
Using an analogous GA-DDH distinguisher to the one depicted in Fig.~\ref{fig:SUFCMADDH}, it is straightforward to verify that for any $\mathsf{PPT}$ adversary \(\A\) running in time \(T\), there are $\mathsf{PPT}$ adversaries \(\B_1\) and \(\B_2\), each running in time \(T + \poly\), such that
\begin{align*}
\Adv^{\pcgame[0]}(\A) &\leq \Adv^{\pcgame[1]}(\A) + \Adv^{\mathrm{GA\text{-}DDH}}(\B_1)\\
\Adv^{\pcgame[1]}(\A) &\leq \Adv^{\pcgame[2]}(\A) + \Adv^{\mathrm{GA\text{-}DDH}}(\B_2).
\end{align*}
Taking \(\B\) to be whichever of \(\B_1\) and \(\B_2\) has greater advantage yields the claim.
\end{proof}
\end{claim}

\begin{claim}
For any algorithm \(\A\) that makes at most \(q\) total queries to \(\H, \OSign, \OSimul\), and \(\OVerify\), we have
\[
    \Adv^{\pcgame[2]}(\A) \leq \Adv^{\pcgame[3]}(\A) + \frac{(q+1)^2}{2^{2\lambda}}.
\]

\begin{proof}
As in the proof of~\Cref{thm:SUFCMA}, this follows by the birthday bound, noting that at most \(q+1\) hash values are defined: \(q\) by the adversary's queries, and one by the challenge signature.
\end{proof}
\end{claim}

We now proceed to \(\pcgame[4]\), depicted in Fig.~\ref{fig:PSIG4}.

\begin{figure}[ht!]
\centering
\begin{pchstack}[boxed]
\procedure[linenumbering, lnstart=100]
{
    $\pcgame[4](1^\lambda)$
}
{
    \ppar \gets \Setup(1^\secpar) \\
    \pcfor i = 0,1\\
    \t (\sk_{S,i}, \pk_{S,i}) \gets \SKgen(\ppar)\\
    (\sk_{V}, \pk_{V}) \gets \VKgen(\ppar)\\
    b  \sample \{0,1\} \\
    \gamechange[white]{$r_0 \sample \ZZ_N$} \\
    \gamechange[white]{$R_0 \gets [\g^{r_0}] \ast E$}\\
    \gamechange[white]{$r_1 \sample \ZZ_N$} \\
    \gamechange[white]{$R_1 \gets [\g^{r_1}] \ast E$}\\
    m^* \gets \adv_1^{\H,\Ocal_{\Simul}, \Ocal_{\Verify}}(\pk_{S,0}, \pk_{S,1}, \pk_V)\\
    \sigma^* := (h^*, z^*) \sample \{0,1\}^{2\lambda} \times \ZZ_N \\
    \gamechange[white]{$\pcassert \nexists x \neq x' \mbox{ s.t. } L[x] = L[x'] \neq \bot$}\\
    b' \gets \adv_2^{\H,\Ocal_{\Simul}, \Ocal'_{\Verify}}(\pk_{S,0}, \pk_{S,1}, \pk_V, m^*, \sigma^*)\\
    \pcreturn [b = b']
}
\begin{pcvstack}[space=1em]
\procedure[lnstart=200,linenumbering]{$\OSimul(i,m)$}
{
    z \sample \ZZ_N \\
    h \sample \{0,1\}^{2\lambda}\\
    \sigma \gets  (h,z) \\
    \Q \gets \Q \cup \{(i,m,\sigma)\} \\
    \pcreturn \sigma
}
\procedure[linenumbering, lnstart=300]{$\H(x)$}
{
    \pcif L[x] = \bot \\
    \t L[x] \sample \{0,1\}^{2\secpar} \\
    \pcreturn L[x]
}
\procedure[lnstart=400,linenumbering]{$\OVerify(i, (m,\sigma))$}
{
    \pcreturn [(i,m,\sigma) \in \Q]
}
\procedure[lnstart=500,linenumbering]{$\OVerify'(i, (m,\sigma))$}
{
    \pcassert (m,\sigma) \neq (m^*, \sigma^*) \\
    \pcreturn [(i,m,\sigma) \in \Q]
}
\end{pcvstack}
\end{pchstack}
\caption{The \(\pcgame[4]\) used in the proof of~\Cref{thm:PSI}.}
\label{fig:PSIG4}
\end{figure}

\begin{claim}
For any algorithm \(\A\) that makes at most \(q_\H\) (respectively, \(q_\Simul, q_\Verify\)) queries to \(\H\) (respectively \(\OSimul, \OVerify\)), we have 
\begin{align*}
\Adv^{\pcgame[3]}(\A) \leq \Adv^{\pcgame[4]}&(\A) + \frac{q_\H q_\Verify}{2^{2\lambda}} + \frac{q_\Simul q_\Verify}{2^{2\lambda} N} \\
&+ \frac{(q_\H + q_\Simul)(2q_\Simul + q_\Verify) + 2(q_\H + q_\Simul + q_\Verify)}{N}. 
\end{align*}

\begin{proof}
From the adversary's perspective, \(\pcgame[3]\) is identical to \(\pcgame[4]\) unless the result of an oracle query is contradicted by a later oracle query, or the challenge message/signature pair \( (m^*, \sigma^*)\) contradicts or is later contradicted by the output of an oracle query. 

We first enumerate the ways that the oracles can contradict one another:

\begin{description}[labelindent=0pt]
\item[\(\OSimul\) contradicts \(\OSimul\):] The adversary queries \(\OSimul\) twice at the same pair \( (i,m)\), and receives two signatures with the same value of \(z\) and different values of \(h\). Or, the adversary makes queries of the form \( \OSimul(0,m)\) and \(\OSimul(1,m)\) and receives \(\sigma_0 = (h_0, z_0)\) and \(\sigma_1 = (h_1, z_1)\), respectively, with \(h_1 \neq h_0\) and \(z_1 = z_0 + r_0 - r_1\).

Each query to \(\OSimul\) uses a fresh, uniformly-random value of \(z\). Thus, the probability that two calls to \(\OSimul\) use the same value of \(z\) is at most \(\frac{q_\Simul^2}{N}\) by the birthday bound. Similarly, the probability that two calls to \(\OSimul\) use values of \(z\) that differ by \(r_0 - r_1\) is at most \(\frac{q_\Simul^2}{N}\), again by the birthday bound. Thus, this event occurs with probability at most \(\frac{2q_\Simul^2}{N}\).
\item[\(\OVerify/\OVerify'\) contradicts \(\OSimul\):] \(\OVerify(i, (m,\sigma))\) returns 0 after \(\OSimul(i,m)\) has returned \(\sigma\). Or, \(\OVerify(i, (m, \sigma_i = (h,z_i))\) returns 0 after \(\OSimul(1-i, m)\) has returned \(\sigma_{1-i} = (h,z_{1-i})\), where \(z_{1-i} = z_i + r_i - r_{1-i}\).

By construction, the first possibility never happens. For the second possibility, until such a query occurs, the value of \(r_0 - r_1\) is information-theoretically hidden from the adversary, and so such a query can only be made by chance. Since there are \(N\) possible values of \(r_0 - r_1\) and the adversary makes at most \(q_\Verify\) queries to \(\OVerify\) and \(\OVerify'\) together and at most \(q_\Simul\) queries to \(\OSimul\), this occurs with probability at most \(\frac{q_\Verify q_\Simul}{N}\).

\item[\(\H\) contradicts \(\OSimul\):] The adversary queries \(\OSimul\) at \( (i,m)\), and it returns a signature \(\sigma = (h,z)\). Later, the adversary queries \(\H\) at \( ([g^z] \ast R_i \,\|\,m)\) and it returns \(h' \neq h\).

Up until such a query occurs, \(R_0\) and \(R_1\) are information-theoretically hidden from the adversary, and so such a query cannot be made except by random chance. There are \(N\) possible values of \(R_0\) (and \(R_1\)), and the adversary makes at most \(q_\H\) queries to \(\H\) and at most \( q_\Simul\) queries to \(\OSimul\). Thus, this event occurs with probability at most \( \frac{q_Hq_\Simul}{N}\).

\item[\(\OSimul\) contradicts \(\OVerify\):] The adversary queries \(\OVerify\) at some \( (i, (m,\sigma = (h, z_i)))\) and it is rejected. Later, the adversary queries \(\OSimul\) at \( (i,m) \) and the oracle returns \( (h, z_i)\), or at \( (1-i, m)\) and it returns \( (h, z_i + r_i - r_{1-i})\).

Since each call to \(\OSimul\) generates independent, uniformly random values of \(z\) and \(h\), and there are at most \(q_\Simul\) queries to \(\OSimul\) and \(q_\Verify\) queries to \(\OVerify\), this event happens with probability at most \(\frac{q_\Verify q_\Simul}{2^{2\lambda} N}\).

\item[\(\OVerify\) contradicts \(\OVerify\):] The adversary queries \(\OVerify\) at some \( (i, (m, \sigma))\) twice, and one query returns 0 while the other returns 1. Or, the adversary queries \(\OVerify(0, (m, \sigma_0 = (h, z_0)))\) and \(\OVerify(1, (m, \sigma_1 = (h, z_1)))\) with \(z_1 = z_0 + r_0 - r_1\), and one query returns 0 while the other returns 1.

The first possibility can only happen in the following way: \( (i, m, \sigma)\) is rejected because it is not in \(\Q\), then a query to \(\OSimul\) adds \( (i, m, \sigma)\) to \(\Q\), and then the adversary queries \(\OVerify\) at \( (i, m, \sigma)\) again. This case is subsumed by the previous one.

The second possibility can only happen in two ways: 
\begin{enumerate}
\item For some \(i \in \{0,1\}\), \( (i, (m, \sigma_i))\) is rejected because it is not in \(\Q\), then a query to \(\OSimul\) adds \( (1-i, m, \sigma_{1-i})\) to \(\Q\), and then the adversary queries \(\OVerify\) at \( (1-i, (m, \sigma_{1-i}))\). This case is subsumed by the previous one.
\item For some \(i \in \{0,1\}\), a query to \(\OSimul\) adds \( (1-i,m, \sigma_{1-i})\) to \(\Q\), and then the adversary queries \(\OSimul\) at both \( (1-i,(m,\sigma_{1-i}))\) and \( (i, (m,\sigma_i))\) (in some order) \emph{without} querying \(\OSimul(i,m)\) and receiving \(\sigma_i\). This is subsumed by the earlier case ``\(\OVerify\) contradicts \(\OSimul\)''.
\end{enumerate}
Altogether, this case is subsumed entirely by earlier ones.

\item[\(\H\) contradicts \(\OVerify\):] The adversary queries \(\OVerify\) at some \( (i, (m, \sigma = (h,z)))\) and it is rejected. Later, the adversary queries \(\H\) at \( ([\g^z] \ast R_i \,\|\,m)\) and it returns \(h\).

Since \(\H([\g^z] \ast R_i \cat m)\) is set uniformly at random when the query to \(\H\) is made, and the adversary makes at most \(q_\H\) queries to \(\H\) and \(q_\Verify\) queries to \(\OVerify\), this occurs with probability at most \(\frac{q_\H q_\Verify}{2^{2\lambda}}\).

\item[\(\OSimul\) contradicts \(\H\):] The adversary queries \(\H(Z \,\|\, m)\) for some curve \(Z\), and it returns \(h\). Later, the adversary queries \(\OSimul\) at \((i,m)\) and the oracle returns \( \sigma = (h',z)\) with \(h' \neq h\) and \([g^z] \ast R_i = Z\).

Since the action used in CSI-FiSh is free, for each \( (i,Z)\) there is exactly one value of \(z\) such that \(Z = [g^z] \ast R_i\). Since each call to \(\OSimul\) generates an independent, uniformly random value of \(z\), and there are at most \(q_\H\) queries to \(\H\) and \(q_\Simul\) queries to \(\OSimul\), this event happens with probability at most \(\frac{q_\H q_\Simul}{N}\).

\item[\(\OVerify\) contradicts \(\H\):] The adversary queries \(\H\) at some \( (Z \,\|\,m)\) and it returns \(h\). Later, the adversary queries \(\OVerify\) at \( (i, (m, \sigma = (h,z)))\) such that \(Z = [g^z] \ast R_i\) and the oracle returns 0.

Up until such a query occurs, \(R_0\) and \(R_1\) are information-theoretically hidden from the adversary, and so such a query cannot be made except by random chance. There are \(N\) possible values of each \(R_i\), and the adversary makes at most \(q_\H\) queries to \(\H\) and at most \( q_\Verify\) queries to \(\OVerify\). Thus, this event occurs with probability at most \( \frac{q_H q_\Verify}{N}\).

\item[\(\H\) contradicts \(\H\):] The adversary queries \(\H\) twice at the same input \(x\) and receives two different outputs. By construction, this never happens.
\end{description}

Now we must consider the ways that the challenge message/signature pair \( (m^*, \sigma^*)\) can contradict or be contradicted by an oracle query. To begin, \( (m^*, \sigma^*)\) implicitly defines exactly one hash value: \(\H([g^{z^*}] \ast R_b \,\|\,m) = h^*\). Since \(z^*\) is chosen uniformly at random when the challenge signature is constructed, \([\g^{z^*}] \ast R_b\) is a uniformly random curve; thus, the probability that this contradicts an existing hash value is at most \(\frac{q_H + q_\Simul + q_\Verify}{N}\).

Later oracle queries may contradict the challenge message/signature pair as follows:
\begin{description}[labelindent=0pt]
\item[\(\OSimul\) contradicts \( (m^*, \sigma^*)\):] The adversary queries \(\OSimul\) at some \( (i, m^*)\), and it returns \(\sigma = (h^*, z)\)  where \( [\g^{z}] \ast R_i = [\g^{z^*}] \ast R_b\).

Since the group action in CSI-FiSh is free, for each \(i\) there is exactly one value of \(z\) such that \( [\g^{z}] \ast R_i = [\g^{z^*}] \ast R_b\). Since \(z\) is chosen uniformly at random in each call to \(\OSimul\), this event occurs with probability at most \(\frac{q_\Simul}{N}\).

\item[\(\OVerify\) contradicts \( (m^*, \sigma^*)\):] The adversary queries \(\OVerify\) at \( (1-b, (m^*, \sigma = (h^*, z)))\), where \( z = z^* + r_b - r_{1-b}\), and the oracle returns 0.

Until this \(\OVerify\) query is made, \(r_b - r_{1-b}\) is information-theoretically hidden from the adversary. Thus, such a query can be made only by chance. Thus, the probability that such a query is made is at most \(\frac{q_\Verify}{N}\).

\item[\(\H\) contradicts \( (m^*, \sigma^*)\):] The adversary queries \(\H\) at \( ([g^{z^*}] \ast R_b \,\|\,m)\) and it returns \(h \neq h^*\).

Until this \(\H\) query is made, \(R_b\) is information-theoretically hidden from the adversary. Thus, such a query can be made only by chance. Thus, the probability that such a query is made is at most \(\frac{q_\H}{N}\).

\end{description}

Summing all of the individual probabilities, we find that
\begin{align*}
\Adv^{\pcgame[3]}(\A) \leq \Adv^{\pcgame[4]}&(\A) + \frac{q_\H q_\Verify}{2^{2\lambda}} + \frac{q_\Simul q_\Verify}{2^{2\lambda} N} \\
&+ \frac{(q_\H + q_\Simul)(2q_\Simul + q_\Verify) + 2(q_\H + q_\Simul + q_\Verify)}{N}.
\end{align*}

\end{proof}
\end{claim}

\begin{claim}
For any adversary \(\A\), we have \(\Adv^{\pcgame[4]}(\A) = 0\).

\begin{proof}
In \(\pcgame[4]\), the challenge signature and all oracle responses are independent of the value of \(b\) chosen by the adversary. Since \(b\) is chosen uniformly at random from \(\{0,1\}\), regardless of how \(\A\) chooses \(b'\), the probability that \(b' = b\) is \(\frac{1}{2}\), and \(\A\)'s advantage is 0.
\end{proof}
\end{claim}

\begin{proof}[Proof (of~\Cref{thm:PSI})]
Altogether, for any adversary \(\A\) running in time \(T\) and making at most \(q_\H\) (respectively, \(q_\Simul, q_\Verify\)) queries to \(\H\) (respectively \(\OSimul, \OVerify\)), there is an adversary \(\B\) that runs in time \(T + \poly[\lambda]\) such that
\begin{align*}
   \Adv^{\PSI}_{\prot}(\A)  &\leq  2\Adv^{\mathrm{GA\text{-}DDH}}(\B) + \frac{q^2 +q_\H q_\Verify}{2^{2\lambda}} + \frac{q_\Simul q_\Verify}{2^{2\lambda} N} \\
    &+ \frac{(q_\H + q_\Simul)(2q_\Simul + q_\Verify) + 2(q_\H + q_\Simul + q_\Verify)}{N}
\end{align*}
where \(q = q_\H + q_\Simul + q_\Verify\). Thus, when \(\A\) is $\mathsf{PPT}$ (so that \(T = \poly[\lambda]\) and \(q = \poly[\lambda]\)), \(\Adv^{\PSI}_{\prot}(\A) = \negl[\lambda]\) under the GA-DDH assumption, as required.
\end{proof}


\section{Tight Reductions to GAIP in the AGAM}
\label{sec:AGAMProofs}

As shown in~\Cref{sec:CSI-SDVS-security}, our security proofs rely on the
GA-DDH assumption. It is desirable to base such reductions on computational
problems, such as GAIP or GA-CDH, as in~\cite{khuc2025}. In this section, we
show that, in the AGAM (see~\Cref{sec:AGAM}) and the random oracle model, the
SUF-CMA and PSI security of \textsf{CSI-SDVS} reduce tightly to the hardness
of GAIP. We first establish SUF-CMA security.

\begin{theorem}
\label{thm:AGAM_SUFCMA}
Assuming the hardness of GAIP (Definition~\ref{def:GAIP}), the proposed scheme, $\prot$, is SUF-CMA secure, as defined in~\Cref{def:SUF-SDVS}, in the
AGAM and the random oracle model.
\end{theorem}

At a high level, the idea of the proof is as follows: we embed the GAIP instance \(X\) into the game either by setting \(\pk_S := X\) or \(\pk_V := X\). In either case, letting \(\A\) denote an adversary in the AGAM and ROM who wins the SUF-CMA game with non-negligible probability, we examine the message/signature pair \( (m, \sigma = (h,z))\) that it produces. With high probability, we will have \(h = \H(Y \,\|\, m)\) for a unique \(Y \in \mathcal{E}\ell\ell_{\mathbb{F}_{p}}(\mathfrak{O})\). Then, because we work in the AGAM, \(\A\) will have provided us with a explanation \( Y = [\g^a] \ast A\) for \(Y\), where \(A\) is among the curves we provided to \(\A\) initially; in particular, \(A \in \{E, \pk_S, \pk_V\}\). Then, if \(A \neq X\), we can extract a solution to the GAIP instance. Since \(X\) is embedded uniformly at random into \(\pk_S\) or \(\pk_V\), and since the reductions behave identically from the perspective of \(\A\) in those two cases, we will be able to extract the GAIP instance solution with non-negligible probability.

The reduction is depicted in Figs.~\ref{fig:Reduction}, \ref{fig:pkSReduction}, and \ref{fig:pkVReduction}.
\begin{figure}[h!]

\centering
\begin{pcvstack}[boxed]

\procedure[linenumbering, lnstart=100]{$\B(X)$}
{
    \ID \sample \{S,V\} \\
    x \gets \B^{\ID}(X) \\
    \pcreturn x
}
\end{pcvstack}
\caption{The reduction \(\B\) from GAIP to SUF-CMA security of \(\prot\).}
\label{fig:Reduction}
\end{figure}

\begin{figure}[p!]
\centering

\begin{pcvstack}[boxed,space=1em]

\begin{pchstack}
\procedure[linenumbering, lnstart=100]{$\B^{S}(X)$}
{
    \Q \gets \varnothing, L \gets [~] \\
    (\sk_V, \pk_V) \gets \prot.\VKgen(\ppar) \\
    \pk_S \gets X \\
    \oracles \gets (\H,\OSign,\OSimul,\OVerify) \\
    (m^*, \sigma^* = (h^*,z^*)) \gets \A^{\oracles}(\pk_S,\pk_V) \\
    \pcassert (m^*,\sigma^*) \not\in \Q  \label{ln:Forgery}  \\
    \pcassert \exists!~Y \in \mathcal{E}\ell\ell_{\mathbb{F}_{p}}(\mathfrak{O}) \mbox{ s.t. }  h^* = L[Y \,\|\, m^*] \label{ln:HashCheck} \\
    \pcassert \prot.\Verify(\sk_V,\pk_S, m^*,\sigma^*) \label{ln:Valid} \\
    \pcassert \GetRep(Y) \neq \bot \label{ln:NoBot} \\
    (a, A) \gets \GetRep(Y) \label{ln:GetRef} \\
    \pcif A = \pk_S : \pcreturn \bot \label{ln:SS} \\
    \pcif A = E : \pcreturn a - z^* - \sk_V \label{ln:SE}\\
    \pcif A = \pk_V : \pcreturn a - z^* \label{ln:SV}
}
\begin{pcvstack}[space=1em]
\procedure[lnstart=200,linenumbering]{$\OSign(m),\,\OSimul(m)$}
{
    z \sample \ZZ_N \\
    h \gets \H([g^{z+\sk_V}] \ast \pk_{S} \,\|\, m)\\
    \sigma \gets  (h,z) \\
    \Q \gets \Q \cup \{(m,\sigma)\}\\
    \pcreturn \sigma
}
\procedure[linenumbering, lnstart=300]{$\H(x)$}
{
    \pcif L[x] = \bot \\
    \t L[x] \sample \{0,1\}^{\secpar} \\
    \pcreturn L[x]
}

\procedure[lnstart=400,linenumbering]{$\OVerify(i,m,\sigma)$}
{
    \pcparse (h,z) \gets \sigma \\
    h' \gets \H([\g^{z+\sk_V}] \ast \pk_{S} \,\|\, m) \\
    \pcreturn [h = h']
}

\end{pcvstack}
\end{pchstack}
\end{pcvstack}

\caption{The reduction when the GAIP instance in embedded in \(\pk_S\).
\label{fig:pkSReduction}
}

\begin{pcvstack}[boxed,space=1em]

\begin{pchstack}
\procedure[linenumbering, lnstart=100]{$\B^{V}(X)$}
{
    \Q \gets \varnothing, L \gets [~] \\
    (\sk_S, \pk_S) \gets \prot.\VKgen(\ppar) \\
    \pk_V \gets X \\
    \oracles \gets (\H,\OSign,\OSimul,\OVerify) \\
    (m^*, \sigma^* = (h^*,z^*)) \gets \A^{\oracles}(\pk_S,\pk_V) \\
    \pcassert (m^*,\sigma^*) \not\in \Q \\
    \pcassert \exists!~Y \in \mathcal{E}\ell\ell_{\mathbb{F}_{p}}(\mathfrak{O}) \mbox{ s.t. }  h = L[Y \,\|\, m^*] \\
    \pcassert \prot.\Verify'(\sk_V,\pk_S, m^*,\sigma^*) \\
    (a, A) \gets \GetRep(Y) \\
    \pcif A = \pk_S : \pcreturn a - z^* \label{ln:VS} \\
    \pcif A = E : \pcreturn a - z^* - \sk_S \label{ln:VE} \\
    \pcif A = \pk_V : \pcreturn \bot \label{ln:VV}
}

\begin{pcvstack}[space=1em]
\procedure[lnstart=200,linenumbering]{$\OSign(m),\,\OSimul(m)$}
{
    z \sample \ZZ_N \\
    h \gets \H([g^{z+\sk_S}] \ast \pk_{V} \,\|\, m)\\
    \sigma \gets  (h,z) \\
    \Q \gets \Q \cup \{(m,\sigma)\}\\
    \pcreturn \sigma
}
\procedure[linenumbering, lnstart=300]{$\H(x)$}
{
    \pcif L[x] = \bot \\
    \t L[x] \sample \{0,1\}^{\secpar} \\
    \pcreturn L[x]
}

\procedure[lnstart=400,linenumbering]{$\OVerify(i,m,\sigma)$}
{
    \pcparse (h,z) \gets \sigma \\
    h' \gets \H([\g^{z+\sk_S}] \ast \pk_{V} \,\|\, m) \\
    \pcreturn [h = h']
}

\end{pcvstack}

\end{pchstack}
\end{pcvstack}

\caption{The reduction when the GAIP instance in embedded in \(\pk_V\).
\label{fig:pkVReduction}
}
\end{figure}

For the proof of~\Cref{thm:AGAM_SUFCMA}, we proceed through a sequence of claims.

\begin{claim}
Let \(\ppar = (p, \g, N, E)\) be the pubic parameters of a GAIP instance \(X\). If \(\bot \neq x \gets \B(X)\), then \( [\g^x] \ast E = X\); that is, \(x\) is the solution to the GAIP instance.

\begin{proof} To begin, we address the case that \(\ID = S\). We note that in this case, \(\B\) will only return \(x \neq \bot\) as a result of lines~\ref{ln:SE} and~\ref{ln:SV} of~\Cref{fig:pkSReduction}. Suppose that \(x\) is returned on line~\ref{ln:SE}, so that \(x = a - z - \sk_V\). Then, since all 3 assert statements of \(\B^S\) have passed, we know that there is a unique \(Y \in \mathcal{E}\ell\ell_{\mathbb{F}_{p}}(\mathfrak{O})\) such that
\[
    \H(Y \,\|\, m) = h^* = \H([\g^{z^* + \sk_V}] \ast \pk_S \,\|\, m).
\]
Since hash values are only defined by queries to \(\H\) (including those made during signing, simulating, and verification), we necessarily have \(Y = [\g^{z^* + \sk_V}] \ast \pk_S\). Moreover, since we are working in the AGAM, we can also construct from \(\A\)'s queries a pair \( (a,A)\) such that \(a \in \ZZ_N\), \(A \in \{\pk_S, \pk_V, E\}\) and \(Y = [\g^a] \ast A\). If \(A = E\), by line~\ref{ln:SE} we have \(x = a - z^* - \sk_V\), and moreover we have (recalling that \(\pk_S = X\) in this case):
\[
    [\g^{z^* + \sk_V}] \ast \pk_S = [\g^a] \ast E \implies X = [\g^{a-z^* - \sk_V}] \ast E = [\g^x] \ast E 
\]
so that, indeed, \(x\) is the solution to the GAIP instance \(X\). Similarly, if \(A = \pk_V\), we have \(x = a -z^*\), and then
\[
    [\g^{z^* + \sk_V}] \ast \pk_S = [\g^a] \ast \pk_V \implies X = [\g^{a - z^*}] \ast E = [\g^x] \ast E 
\]
(where we have used the identity \(\pk_V = [\g^{\sk_V}] \ast E\)) so that, again, \(x\) is a solution to the GAIP instance \(X\).

The argument when \(\ID = V\) is similar, and so we omit it. 
\end{proof}
\end{claim}

\begin{claim}
\(\B^S\) and \(\B^V\) perfectly simulate the SUF-CMA game to \(\A\). In particular: fix public parameters \( (p, [\g], N, E, \H)\) for \(\prot\). Let \(\mathrm{View}(\A)\) denote the adversary's \emph{view} in the SUF-CMA game---more precisely, \(\mathrm{View}(\A)\) is the joint distribution of the public values \( \pk_S, \pk_V\) and all oracle responses given to \(\A\) in the course of the SUF-CMA experiment. Similarly, let \(\mathrm{View}_S(\A)\) and \(\mathrm{View}_V(\A)\) denote the adversary's view in the reductions \(\B^S\) and \(\B^V\), respectively, when \(X\) is a uniformly-chosen GAIP instance. Then, the distributions \(\mathrm{View}(\A), \mathrm{View}_S(\A)\), and \(\mathrm{View}_V(\A)\) are identical.
\end{claim}

\begin{proof}
We prove that the distributions \(\mathrm{View}(\A),\) and \(\mathrm{View}_S(\A)\) are identical; the proof for \(\mathrm{View}_V(\A)\) is similar.

To begin, note that the public values \( (\pk_S, \pk_V)\) are distributed uniformly from \(\mathcal{E}\ell\ell_{\mathbb{F}_{p}}(\mathfrak{O})^2\) in both games, as required. We then proceed to consider oracle queries. Note that \(\OSimul, \H\), and \(\OVerify\) behave identically in both games, so we need only consider \(\OSign\). Note that in \(\B^S\), \(\OSign(m)\) actually runs \(\prot.\Simul(m)\); by Theorem~\ref{thm:Identical}, this also is exactly the same behaviour as \(\prot.\Sign\), which behaves identically to \(\OSign\) in the SUF-CMA game. Thus, indeed, the adversary's view is the same in both the true SUF-CMA game and in \(\B^S\).

The proof for \(\B^V\) is analogous.
\end{proof}

Consequently, the probability that \(\A\)'s response to \(\B\) is one that would win the SUF-CMA game is exactly \(\Adv^{\SUFCMA,\A}_{\prot}(\secpar)\).

\begin{claim} Let \(\A\) be an adversary for the SUF-CMA game against \(\prot\). Suppose that \(\A\) makes at most \(q\) total queries to the oracles \(\H, \OSign, \OSimul\), and  \(\OVerify\). Define
\[
    \arraycolsep1.6pt
    \Succ^{\B^\A}(\secpar) = \PP\left[ x' \neq \bot \middle| \begin{array}{rcl}(p,\g,N,E,\H) &\gets& \Setup(1^\secpar) \\ x &\sample& \ZZ_N \\ X &\gets& [\g^x] \ast E \\ x' &\gets& \B^\A(X) \end{array} \right]. 
\]
Then 
\[
    \Succ^{\B^\A}(\secpar) \geq \frac{1}{2}\Adv^{\SUFCMA,\A}_{\prot}(\secpar) - \frac{q^2 + q_\Verify + 1}{2^{2\secpar}}. 
\]

\begin{proof} 
Since \(\ID\) is sampled uniformly at random from \(\{S,V\}\) in the execution of \(\B^\A\), we have
\[
    \Succ^{\B^\A}(1^\secpar) = \frac{1}{2}\PP[\B^{\A}(X) \neq \bot \,|\, \ID = S] +\frac{1}{2}\PP[\B^{\A}(X) \neq \bot \,|\, \ID = V] 
\]
and so it suffices to analyze the success probability of \(\B^S\) and \(\B^V\) individually. Beginning with \(\B^S\), referring to Fig.~\ref{fig:pkSReduction}, it is clear that \(\B^S\) succeeds as long as:
\begin{enumerate}\itemsep-0.0ex
\item \(\A^{\oracles}(\pk_S,\pk_V)\) wins the instance of the  SUF-CMA game;
\item When \(A\) is constructed on line~\ref{ln:GetRef}, we have \(A \neq \pk_S\);
\item The assertion on line~\ref{ln:HashCheck} does not fail; and,
\item The assertion on line~\ref{ln:NoBot} does not fail.
\end{enumerate}
On inputs of length \(\secpar\), the first condition is satisfied with probability \(\Adv^{\SUFCMA,\A}_{\prot}(1^\secpar)\) by definition. From now on, assume that \(\A\) has won the SUF-CMA game.

The cases on lines~\ref{ln:SS}--\ref{ln:SV} are exhaustive; let \(p_S, p_E, p_V\) denote the probability that---in addition to \(\A\) succeeding in the SUF-CMA game---\(A = \pk_S\), \(A=E\), and \(A=\pk_V\), respectively. We have 
\[
    \Adv^{\SUFCMA,\A}_{\prot}(1^\secpar) = p_S + p_E + p_V.
\]
When computing \(\PP[\B^\A(X) \neq \bot \mid \ID = S]\), we must subtract \(p_S\) from this probability, because if \(A = \pk_S\) then \(\B^S\) fails on line~\ref{ln:SS}. We must also subtract the probability of the assertion on line~\ref{ln:HashCheck} fails. There are two ways that it can fail:
\begin{enumerate}
\item There does not exist any \(Y \in \mathcal{E}\ell\ell_{\mathbb{F}_{p}}(\mathfrak{O})\) such that \(h^* = L[Y \,\|\, m^*]\); or,
\item There exist two or more \(Y \in \mathcal{E}\ell\ell_{\mathbb{F}_{p}}(\mathfrak{O})\) such that \(h^* = L[Y \,\|\, m^*]\).
\end{enumerate}
In the first case, \(\A\)'s signature verifies despite the fact that \(\A\) has not queried (directly or indirectly) \( \H(Y \,\|\, m^*)\); this occurs with probability \({2^{-2\secpar}}\), since \(\H(Y \,\|\, m^*)\) would be chosen uniformly at random while verifying \(\A\)'s forgery during the SUF-CMA game. In the second case, there has been a collision in \(\H\), which occurs with probability at most \( \frac{q^2}{2^{2\secpar}}\) by the birthday bound.

We must also subtract the probability that the assertion on line~\ref{ln:NoBot} fails. Note that we only reach this line if the assertion on line~\ref{ln:HashCheck} succeeds, so we may assume that there is some \(Y \in \mathcal{E}\ell\ell_{\mathbb{F}_{p}}(\mathfrak{O})\) such that \(h^* = L[Y \,\|\, m^*]\). Moreover, the assertion on line \ref{ln:Forgery} must also have passed, meaning that \(L[Y \,\|\, m^*]\) was not defined by a query to \(\OSign\) or \(\OSimul\). Thus, the only way for the assertion on line~\ref{ln:NoBot} to fail is if \(L[Y \,\|\, m^*]\) is defined by a query to \(\OVerify\), and its value is never revealed by \(\OSign, \OSimul\), or \(\H\). If that is the case, the probability that the assertion on line~\ref{ln:Valid} passes is at most \(\frac{q_\Verify+1}{2^{2\lambda}}\), since each of the \(q_\Verify\) queries to \(\OVerify\) can eliminate at most one incorrect value of \(h \neq h^*\).

It follows that the probability that \(\B^S\) succeeds is at least
\[
    \PP[\B^\A(1^\secpar) \neq \bot~|~\ID = S] \geq p_E + p_V - \frac{q^2 + q_\Verify + 1}{2^{2\secpar}}. 
\]
A similar argument yields 
\[
    \PP[\B^\A(1^\secpar) \neq \bot~|~\ID = V] \geq p_E + p_S - \frac{q^2 + q_\Verify + 1}{2^{2\secpar}} 
\]
where we note that, from the adversary's perspective, the reductions \(\B^S\) and \(\B^V\) are identical, so the probabilities that \(A = E, \pk_S\), or \(\pk_V\) are the same in both reductions. Thus,
\[
    \Succ^{\B^{\A}}(1^\secpar) \geq p_E + \frac{1}{2}p_S + \frac{1}{2}p_V - \frac{q^2 + 1}{2^{2\secpar}} \geq \frac{1}{2}\Adv^{\SUFCMA,\A}_{\prot}(1^\secpar) - \frac{q^2 + q_\Verify + 1}{2^{2\secpar}}
\]
as required.

Under the GAIP assumption, if \(\A\) is polynomial time then \(\Succ^{\B^\A}(1^\lambda)\) is negligible, as is \(\frac{q^2 + q_\Verify + 1}{2^{2\lambda}}\); thus, \(\Adv^{\SUFCMA,\A}_{\prot}(1^\secpar)\) is negligible, establishing the SUF-CMA security of \(\prot\).
\end{proof}
\end{claim}

\begin{theorem}
\label{thm:AGAM_PSI}
Assuming the hardness of GAIP (Definition~\ref{def:GAIP}), the proposed scheme, $\prot$, is PSI secure, as defined in~\Cref{def:PSI-SDVS}, in the
AGAM and the random oracle model.
\end{theorem}

At a high level, the proof of~\Cref{thm:AGAM_PSI} uses the same technique as that of~\Cref{thm:AGAM_SUFCMA}: any algebraic adversary who wins the PSI game with non-negligible advantage must make a query to the random oracle from which we can extract a solution to a GAIP instance. The main difference arises from how we guarantee that the adversary makes a random oracle query of the correct form.
While in the SUF-CMA game this was guaranteed by the check on line~\ref{ln:Valid} of \(\B^{\ID}\), we need a sequence of games reduction for the analogous statement in the PSI setting. 

The proof proceeds in two phases: first we prove a tight relationship between any adversary's PSI advantage and the probability that they make a query of a specific form related to the challenge signature; then, we prove that we can use any adversary that makes such a query to extract a solution to a GAIP instance.

\begin{lemma}
\label{lem:AGAM_PSI_Queries}
Let \(\A = (\A_1, \A_2)\) be an algebraic adversary for the PSI experiment against \(\prot\) in the ROM that takes time at most \(T\) and makes at most \(q_\H\) queries to \(\H\), \(q_\Simul\) queries to \(\OSimul\), and \(q_\Verify\) queries to \(\OVerify\) and \(\OVerify'\). Then there exists an algorithm \(\C\) for GAIP that runs in time \(T + \poly[\lambda]\) such that  
\begin{align*}
\Adv^{\PSI,\A}_{\prot}(1^\lambda) \leq &~\Adv^{\mathrm{GAIP}, \C}(1^\lambda) + \frac{q_\Simul + q_\Verify}{N} + \frac{q_\Verify}{2^{2\lambda}} \\
&\hspace{18pt} + \PP[\A \mbox{ queries } \H \mbox{ at } ([\g^{z^* + \sk_V}] \ast \pk_{S,b} \,\|\, m^*)]
\end{align*}
where \(b\) is the value chosen on line \ref{ln:fig_PSI_random_coin} of Fig.~\ref{fig:PSI}.

\begin{proof}

\begin{figure}[ht!]
\centering
\begin{pcvstack}[boxed, space=1em]
\begin{pchstack}
\procedure[linenumbering, lnstart=100]
{
    \gamechange[gc1]{$\pcgame[0](1^\lambda)$}
    \gamechange[gc2]{$\pcgame[1](1^\lambda)$}
}
{
    L \gets [~]\\
    \ppar \gets \Setup(1^\secpar) \\
    \pcfor i = 0,1\\
    \t (\sk_{S,i}, \pk_{S,i}) \gets \SKgen(\ppar)\\
    (\sk_{V}, \pk_{V}) \gets \VKgen(\ppar)\\
    b  \sample \{0,1\} \\
    m^* \gets \adv_1^{\H,\Ocal_{\Simul}, \Ocal_{\Verify}}(\pk_{S,0}, \pk_{S,1}, \pk_V)\\
    \gamechange[gc1]{$\sigma^* = (h^*, z^*) \gets \Sign(\sk_{S,b}, \pk_V, m^*)$}\\
    \gamechange[gc2]{$\sigma^* = (h^*, z^*) \sample \{0,1\}^{2\lambda} \times \ZZ_N$} \label{ln:fig_AGAM_PSI_random_sig}\\
    b' \gets \adv_2^{\H,\Ocal_{\Simul}, \Ocal'_{\Verify}}(\pk_{S,0}, \pk_{S,1}, \pk_V, m^*, \sigma^*)\\
    \pcreturn [b = b']
}
\begin{pcvstack}[space=1em]
\procedure[lnstart=200,linenumbering]{$\OSimul(i,m)$}
{
    z \sample \ZZ_N \\
    h \gets \H([g^{z+\sk_V}] \ast \pk_{S,i} \,\|\, m)\\
    \sigma \gets  (h,z) \\
    \pcreturn \sigma
}
\procedure[linenumbering, lnstart=400]{$\H(x)$}
{
    \pcif L[x] = \bot \\
    \t L[x] \sample \{0,1\}^{\secpar} \\
    \pcreturn L[x]
}

\procedure[lnstart=500,linenumbering]{$\OVerify(i,m,\sigma)$
\gamechange{$\OVerify'(i,m,\sigma)$}
}
{
    \gamechange{$\pcassert (m,\sigma) \neq (m^*,\sigma^*)$}\\
    \pcparse (h,z) \gets \sigma \\
    h' \gets \H([\g^{z+\sk_V}] \ast \pk_{S,i} \,\|\, m) \label{ln:fig_AGAM_PSI_ver_h'}\\
    \pcreturn [h = h'] \label{ln:fig_AGAM_PSI_ver_return}
}

\end{pcvstack}
\end{pchstack}

\end{pcvstack}

\caption{Two games used in the proof of Lemma~\ref{lem:AGAM_PSI_Queries}.}
\label{fig:AGAM_PSI}
\end{figure}

Consider the games \(\pcgame[0]\) and \(\pcgame[1]\) depicted in Fig.~\ref{fig:AGAM_PSI}. The two games behave identically from the adversary's perspective unless one of the adversary's queries to \(\H, \OSimul, \OVerify\), or \(\OVerify'\) contradicts the definition of \(\sigma^*\) in \(\pcgame[1]\). In more detail: in \(\pcgame[0]\), the construction of \(\sigma^*\) defines a single random oracle value \(\H([\g^{z^*+\sk_V}] \ast \pk_{S,b} \,\|\, m^*) = h^*\), and this is used consistently by the other oracles. In contrast, in \(\pcgame[1]\), this random oracle value is \emph{not} programmed when \(\sigma^*\) is constructed, meaning that if the adversary queries (directly, or indirectly by a call to \(\OSimul, \OVerify\), or \(\OVerify'\)) \(\H\) at this same input, the two games may differ. So we bound the probability that the adversary (directly or indirectly) queries this hash value in a way that contradicts \(\sigma^*\).

We consider each oracle separately:
\begin{description}
\item[$\H$ contradicts $\sigma^*$:] This can happen only if \(\A\) queries \(\H\) at \( ([g^{z^*+\sk_V}] \ast\pk_{S,b} \,\|\, m^*)\), either before or after \(\sigma^*\) is constructed.

\item[$\OSimul$ contradicts $\sigma^*$:] This happens if and only if $\A$ queries $\OSimul(i,m^*)$, the oracle chooses a value of $z$ such that \([\g^{z+\sk_V}] \ast \pk_{S,i} = [\g^{z^*+\sk_V}] \ast \pk_{S,b}\), and then the random oracle output is programmed as some \(h' \neq h^*\). Since the CSIDH group action is free, for each \(i \in \{0,1\}\), there is a unique value of \(z \in \ZZ_N\) for which this occurs: \(z = z^* + \sk_{S,b} - \sk_{S,i}\). Since \(\OSimul\) chooses each value of \(z\) independently and uniformly at random, this occurs with probability at most \(\frac{q_\Simul}{N}\).

\item[\(\OVerify\) contradicts \(\sigma^*\):] In this case, \(\A\) must have queried \(\OVerify\) at some \( (i,m^*,\sigma = (h,z)) \) such that \([\g^{z+\sk_V}] \ast \pk_{S,i} = [\g^{z^*+\sk_V}] \ast \pk_{S,b}\), before \(\sigma^*\) was constructed. Since \(z^*\) is chosen uniformly at random on line \ref{ln:fig_AGAM_PSI_random_sig} and the group action is free, this happens with probability at most \(\frac{q_\Verify}{N}\).

\item[\(\OVerify'\) contradicts \(\sigma^*\):] A query \( (i, m, \sigma = (h,z)) \) to \(\OVerify'\) can only contradict \(\sigma^*\) if \(m = m^*\) and \([\g^{z+\sk_V}] \ast \pk_{S,i} = [\g^{z^*+\sk_V}] \ast \pk_{S,b}\). Since the group action is free, this means that \(z = z^* + \sk_{S,b} - \sk_{S,i}\). We consider two cases:

\begin{enumerate}
\item If \(i = b\), then \(z = z^*\). The oracle \(\OVerify'\) will not answer the query \( (i, m^*, (h^*, z^*))\), and so we may assume that \(h \neq h^*\). In \(\pcgame[0]\), any such query would return \(0\) because of the failed check on line \ref{ln:fig_AGAM_PSI_ver_return}. Thus, it suffices to bound the probability that this query returns 1. We consider two cases:
\begin{enumerate}
\item A query to \(\H\) or \(\OSimul\) reveals the value of \(L[[\g^{z^*+\sk_V}] \ast \pk_{S,b} \,\|\, m^*]\). This is subsumed by the first two cases.
\item No query to \(\H\) or \(\OSimul\) reveals the value of \(L[[\g^{z^*+\sk_V}] \ast \pk_{S,b} \,\|\, m^*]\). In this case, the value \(h'\) of \(L[[\g^{z^*+\sk_V}] \ast \pk_{S,b} \,\|\, m^*]\) is set uniformly at random on Line \ref{ln:fig_AGAM_PSI_ver_h'}, either by a call to \(\OVerify\) before \(\sigma^*\) is defined, or by a call to \(\OVerify'\) after \(\sigma^*\) is defined. In any case, each query \((b, m^*, \sigma = (h,z^*))\) to \(\OVerify'\) that returns 0 reveals one \emph{incorrect} value of \(L[[\g^{z^*+\sk_V}] \ast \pk_{S,b} \,\|\, m^*]\), and no information about the correct value. After at most \(q_{\Verify}\) such queries, the probability that one of these queries has had the correct value \(h = h'\) (causing the oracle to output 1) is at most \(\frac{q_\Verify}{2^{2\lambda}}\).
\end{enumerate}

\item If \(i = 1 - b\), then \(z = z^* + \sk_{S,b} - \sk_{S,1-b} \). If \(\A\) is an adversary running in time \(T\) that makes such a query with probability \(\varepsilon\), then there exists an algorithm \(\C\) that runs in time \(T + \poly[\lambda]\) which solves the GAIP problem with probability \(\lambda\). In particular, the adversary is given in Fig.~\ref{fig:PSIQueryToGAIP}.
\end{enumerate}
\end{description}

\begin{figure}[h!]
\centering
\begin{pcvstack}[boxed, space=1em]
\begin{pchstack}
\procedure[linenumbering, lnstart=100]
{
    $\C(X_0, X_1)$
}
{
    V \gets \varnothing, L \gets [~]\\
    \ppar \gets \Setup(1^\secpar) \\
    \pcfor i = 0,1\\
    \t \pk_{S,i} \gets X_i\\
    (\sk_{V}, \pk_{V}) \gets \VKgen(\ppar)\\
    b  \sample \{0,1\} \\
    m^* \gets \adv_1^{\H,\Ocal_{\Simul}, \Ocal_{\Verify}}(\pk_{S,0}, \pk_{S,1}, \pk_V)\\
    \gamechange[white]{$\sigma^* = (h^*, z^*) \gets \Simul(\sk_V, \pk_{S,b}, m^*)$}\\
    b' \gets \adv_2^{\H,\Ocal_{\Simul}, \Ocal'_{\Verify}}(\pk_{S,0}, \pk_{S,1}, \pk_V, m^*, \sigma^*)\\
    \pcfor z \in V \\
    \t \pcif~[\g^{z^* - z}] \ast \pk_{S,b} = \pk_{S,1-b}\\
    \t[2] \pcreturn (-1)^b(z^*-z) \\
    \pcreturn \perp
}
\begin{pcvstack}[space=1em]

\procedure[lnstart=200,linenumbering]{$\OSimul(i,m)$}
{
    z \sample \ZZ_N \\
    h \gets \H([g^{z+\sk_V}] \ast \pk_{S,i} \,\|\, m)\\
    \sigma \gets  (h,z) \\
    \pcreturn \sigma
}
\procedure[linenumbering, lnstart=300]{$\H(x)$}
{
    \pcif L[x] = \bot \\
    \t L[x] \sample \{0,1\}^{\secpar} \\
    \pcreturn L[x]
}

\procedure[lnstart=400,linenumbering]{$\OVerify(i,m,\sigma)$
\gamechange{$\OVerify'(i,m,\sigma)$}
}
{
    \gamechange{$\pcassert (m,\sigma) \neq (m^*,\sigma^*)$}\\
    \pcparse (h,z) \gets \sigma \\
    \gamechange{$\pcif (i,m) = (1-b, m^*)$} \label{ln:fig_PSIQueryToGAIP_ver_condition'}\\
    \t \gamechange{$V \gets V \cup \{z\}$}\label{ln:fig_PSIQueryToGAIP_ver_condition''}\\
    h' \gets \H([\g^{z+\sk_V}] \ast \pk_{S,i} \,\|\, m) \\
    \pcreturn [h = h']
}

\end{pcvstack}
\end{pchstack}
\end{pcvstack}
\caption{A GAIP algorithm \(\C\) using a PSI adversary \(\A\) as a subroutine.}
\label{fig:PSIQueryToGAIP}
\end{figure}
\noindent We make three observations about \(\C\):
\begin{enumerate}
\item \(\C\) runs in time \(T + \poly[\lambda]\). This is clear, since \(\A\) runs in time \(T\), and each other part of \(\C\)'s computation runs in time \(\poly[\lambda]\).
\item \(\C\) simulates the PSI game perfectly to \(\A\). We see that the public keys that \(\C\) provides in Fig.~\ref{fig:PSIQueryToGAIP} are distributed indentically to \(\pcgame[0]\) (in particular: they are independent and uniform in \(\mathcal{E}\ell\ell_{\mathbb{F}_{p}}(\mathfrak{O})\)). The oracles behave identically except for the additional bookkeeping on lines \ref{ln:fig_PSIQueryToGAIP_ver_condition'}--\ref{ln:fig_PSIQueryToGAIP_ver_condition''}, which does not affect \(\A\)'s view. And the challenge signature is generated using \(\Simul\) instead of \(\Sign\), but for our protocol \(\prot\) these algorithms' outputs follow identical distributions, so from \(\A\)'s perspective nothing changes. This means that the probability that \(\A\) makes a query of type described in point 2. above is the same when interacting with \(\C\) as it is when playing \(\pcgame[0]\).
\item  If \(\C\) does not output \(\perp\), then it outputs a solution to the GAIP instance \( (X_0, X_1)\). This is straightforward to see: if \(b = 0\) then \(\C\) outputs \(z^* - z\) where \([\g^{z^*-z}] \ast \pk_{S,0} = \pk_{S,1}\), which is precisely a solution to the GAIP instance \( (X_0, X_1)\) (recalling that \(\pk_{S,b} = x_b\) for \(b = 0,1\)). Similarly, if \(b = 1\), then \(\C\) outputs \(z - z^*\), where \([\g^{z^*-z}] \ast \pk_{S,1} = \pk_{S,0}\); equivalently, \([\g^{z-z^*}] \ast \pk_{S,0} = \pk_{S,1}\), so that \(z - z^*\) is the solution to the GAIP instance \( (X_0, X_1)\), as before.
\end{enumerate}
Finally, note that \(\C\) outputs something other than \(\perp\) if and only if \(\A\) makes a query to \(\OVerify'\) of the form \( (1-b, m^*, \sigma = (h,z))\) in which \([\g^{z^* - z}] \ast \pk_{S,b} = \pk_{S,1-b}\); equivalently, in which \(z = z^* + \sk_{S,b} - \sk_{S,1-b}\). This is precisely the type of query whose probability we wished to bound; thus, we have 
\[
\PP[\A \mbox{ queries } \OVerify'(1-b, m^*, \sigma = (h,z)) \mbox{, where } [\g^{z^* - z}] \ast \pk_{S,b} = \pk_{S,1-b}] = \Adv^{\mathrm{GAIP},\C}(1^\lambda)
\]
as required.

Summing the probabilities of all cases, we have that for any algebraic PSI adversary \(\A\) running in time \(T\) and which makes at most \(q_\H\) queries to \(\H\), \(q_\Simul\) queries to \(\OSimul\), and \(q_\Verify\) total queries to \(\OVerify\) and \(\OVerify'\), there exists an algorithm \(\C\) running in time \(T+\poly[\lambda]\) such that
\begin{align*}
\Adv^{\pcgame[0],\A}_{\prot}(1^\lambda) \leq &~\Adv^{\pcgame[1],\A}_{\prot}(1^\lambda) + \Adv^{\mathrm{GAIP}, \C}(1^\lambda) + \frac{q_\Simul + q_\Verify}{N} + \frac{q_\Verify}{2^{2\lambda}} \\
&\hspace{18pt} +\PP[\A \mbox{ queries } \H \mbox{ at } ([\g^{z^* + \sk_V}] \ast \pk_{S,b} \| m^*)].
\end{align*}
Finally, we note that \(\pcgame[0]\) is the PSI game, and that no adversary---even unbounded---can win \(\pcgame[1]\) with probability different from \(\frac{1}{2}\), since \(\sigma^*\) and all oracle outputs are independent of \(b\), meaning that \(\Adv^{\pcgame[1],\A}_{\prot}(1^\lambda) = 0\). Thus,

\begin{align}
\nonumber
\Adv^{\PSI,\A}_{\prot}(1^\lambda) \leq &~\Adv^{\mathrm{GAIP}, \C}(1^\lambda) + \frac{q_\Simul + q_\Verify}{N} + \frac{q_\Verify}{2^{2\lambda}} \\
\label{eqn:PSIQueryProb1}
&\hspace{18pt} + \PP[\A \mbox{ queries } \H \mbox{ at } ([\g^{z^* + \sk_V}] \ast \pk_{S,b} \,\|\, m^*)].
\end{align}

as required.
\end{proof}
\end{lemma}

If \(\A\) runs in polynomial time and we assume the GAIP problem is hard (so that \(\Adv^{\mathrm{GAIP},\C}(1^\lambda) = \negl[\lambda]\)), \Cref{eqn:PSIQueryProb1} implies that
\[
\PP[\A \mbox{ queries } \H \mbox{ at } ([\g^{z^* + \sk_V}] \ast \pk_{S,b} \,\|\, m^*)] \geq \Adv^{\PSI,\A}_{\prot}(1^\lambda) - \negl[\lambda].
\]
Intuitively: an algebraic adversary that breaks the PSI security of $\prot$ with non-negligible advantage makes a query of the form \(\H([\g^{z^* + \sk_V}] \ast \pk_{S,b} \,\|\, m^*)\) non-negligible probability.

We now proceed to prove the second lemma required for our PSI security result.

\newcommand{\E}{\ensuremath{\mathcal{E}}}
\begin{lemma}
\label{lem:QueryImpliesGAIP}
Let \(\A\) be a algebraic adversary for the PSI game against \(\prot\) that runs in time \(T\) and which makes at most \(q_\H\) queries to \(\H\), \(q_\Simul\) queries to \(\Simul\), and \(q_\Verify\) queries to \(\OVerify\) and \(\OVerify'\) together. Then there exists an algorithm \(\E\) for the GAIP problem that runs in time \(T+\poly[\lambda]\), such that
\[
\Adv^{\mathrm{GAIP},\E}(\lambda) \geq \frac{1}{3}\Adv^{\PSI,\A}_{\prot}(\lambda) - \frac{(q+1)^2 + q_\Verify}{3\cdot 2^{2\lambda}} - \frac{q_\Simul + q_\Verify}{3N}.
\]
where \(q = q_\H + q_\Simul + q_\Verify\).
\end{lemma}

\begin{proof}
Let \(\A\) be PSI adversary as described in the Lemma. We define a GAIP adversary \(\D\) in Figs.~\ref{fig:PSIReduction}, \ref{fig:PSIReductionS} and~\ref{fig:PSIReductionV}.
\begin{figure}[h!]

\centering
\begin{pcvstack}[boxed]

\procedure[linenumbering, lnstart=100]{$\D(X)$}
{
    \ID \sample \{S,V\} \\
    x \gets \D^{\ID}(X) \\
    \pcreturn x
}
\end{pcvstack}
\caption{The reduction \(\D\) from GAIP to PSI security of \(\prot\) in the AGAM.}
\label{fig:PSIReduction}
\end{figure}
\end{proof}

\begin{figure}[p!]
\centering
\begin{pchstack}[boxed,space=0.15em]
\procedure[linenumbering, lnstart=100, codesize=\small]{$\D^{S}(X)$}
{
    L \gets [~] \\
    r \sample \ZZ_N \\
    (\pk_{S,0}, \pk_{S,1}) \gets (X, [\g^{r}] \ast X) \\
    (\sk_{V}, \pk_{V}) \gets \prot.\VKgen(\ppar) \\
    (m^*,\st_\A) \gets \A_1^{\H,\OSimul,\OVerify}(\pk_{S,0},\pk_{S,1},\pk_V) \\
    b \sample \{0,1\} \\
    \sigma^* = (h^*,z^*) \gets \prot.\Simul(\sk_V,\pk_{S,b},m^*) \label{ln:fig_PSIReductionS_simul}\\
    b' \gets \A_2^{\H,\OSimul,\OVerify'}(\pk_{S,0}, \pk_{S,1}, \pk_V, \st_\A) \\
    \pcassert \exists!~Y \in \mathcal{E}\ell\ell_{\mathbb{F}_{p}}(\mathfrak{O}) \mbox{ s.t. } h^* = L[Y \,\|\, m^*]  \label{ln:fig_PSIReductionS_assert_1}\\
    \pcassert \GetRep(Y) \neq \bot \label{ln:fig_PSIReductionS_assert_2}\\
    (a,A) \gets \GetRep(Y) \label{ln:fig_PSIReductionS_GetRep}\\
    \pcif A = E : \pcreturn a - z^* - \sk_V - b\cdot r \label{ln:fig_PSIReductionS_if_1}\\
    \pcif A = \pk_V : \pcreturn a - z^* - b\cdot r \label{ln:fig_PSIReductionS_if_2}\\
    \pcreturn \bot
}

\begin{pcvstack}[space=1em]
\procedure[lnstart=200,linenumbering]{$\OSimul(i,m)$}
{
    z \sample \ZZ_N \\
    h \gets \H([g^{z+\sk_V}] \ast \pk_{S,i} \,\|\, m)\\
    \sigma \gets  (h,z) \\
    \pcreturn \sigma
}
\procedure[linenumbering, lnstart=300]{$\H(x)$}
{
    \pcif L[x] = \bot \\
    \t L[x] \sample \{0,1\}^{\secpar} \\
    \pcreturn L[x]
}

\procedure[lnstart=400,linenumbering]{$\OVerify(i,m,\sigma)$
\gamechange{$\OVerify'(i,m,\sigma)$}
}
{
    \gamechange{$\pcassert (m,\sigma) \neq (m^*,\sigma^*)$}\\
    \pcparse (h,z) \gets \sigma \\
    h' \gets \H([\g^{z+\sk_V}] \ast \pk_{S,i} \,\|\, m) \\
    \pcreturn [h = h']
}

\end{pcvstack}
\end{pchstack}
\caption{The reduction from GAIP to PSI security of \(\prot\) when the GAIP instance \(X\) is embedded in \(\pk_{S,0}\) and \(\pk_{S,1}\).
\label{fig:PSIReductionS}
}
\vspace{-7pt}
\begin{pchstack}[boxed,space=0.15em]
\procedure[linenumbering, lnstart=100]{$\D^{V}(X)$}
{
    L \gets [~] \\
    \pcfor i =0,1 \\
    \t (\sk_{S,i}, \pk_{S,i}) \gets \prot.\SKgen(\ppar) \\
    \pk_V \gets X \\
    (m^*,\st_\A) \gets \A_1^{\H,\OSimul,\OVerify}(\pk_{S,0},\pk_{S,1},\pk_V) \\
    b \sample \{0,1\} \\
    \sigma^* = (h^*,z^*) \gets \prot.\Sign(\sk_{S,b},\pk_{V},m^*) \label{ln:fig_PSIReductionV_sign}\\
    b' \gets \A_2^{\H,\OSimul,\OVerify'}(\pk_{S,0}, \pk_{S,1}, \pk_V, \st_\A) \\
    \pcassert \exists!~Y \in \mathcal{E}\ell\ell_{\mathbb{F}_{p}}(\mathfrak{O}) \mbox{ s.t. } h^* = L[Y \,\|\, m^*] \label{ln:fig_PSIReductionV_assert_1}\\
    \pcassert \GetRep(Y) \neq \bot \label{ln:fig_PSIReductionV_assert_2}\\
    (a,A) \gets \GetRep(Y) \\
    \pcif A = E : \pcreturn a - z^* - \sk_{S,b} \label{ln:fig_PSIReductionV_if_1}\\
    \pcif A = \pk_{S,b} : \pcreturn a - z^* \label{ln:fig_PSIReductionV_if_2}\\
    \pcif A = \pk_{S,1-b} : \pcreturn a - z^* - \sk_{S,b} + \sk_{S,1-b} \label{ln:fig_PSIReductionV_if_3}\\
    \pcreturn \bot
}

\begin{pcvstack}[space=1em]
\procedure[lnstart=200,linenumbering]{$\OSimul(i,m)$}
{
    z \sample \ZZ_N \\
    h \gets \H([g^{z+\sk_{S,i}}] \ast \pk_{V} \,\|\, m)\\
    \sigma \gets  (h,z) \\
    \pcreturn \sigma
}
\procedure[linenumbering, lnstart=300]{$\H(x)$}
{
    \pcif L[x] = \bot \\
    \t L[x] \sample \{0,1\}^{\secpar} \\
    \pcreturn L[x]
}

\procedure[lnstart=400,linenumbering]{$\OVerify(i,m,\sigma)$
\gamechange{$\OVerify'(i,m,\sigma)$}
}
{
    \gamechange{$\pcassert (m,\sigma) \neq (m^*,\sigma^*)$}\\
    \pcparse (h,z) \gets \sigma \\
    h' \gets \H([\g^{z+\sk_{S,i}}] \ast \pk_{V} \,\|\, m) \\
    \pcreturn [h = h']
}

\end{pcvstack}
\end{pchstack}
\caption{The reduction from GAIP to PSI security of \(\prot\) when the GAIP instance \(X\) is embedded in \(\pk_V\).
\label{fig:PSIReductionV}
}
\end{figure}

The proof of Lemma~\ref{lem:QueryImpliesGAIP} proceeds in a series of claims.

\begin{claim}
Let \(\ppar = (p,\g,N,E)\) be the public parameters of a GAIP instance \(X\). If \(\bot \neq x \gets \D(X)\), then \([\g^x] \ast E = X\); that is, \(x\) is the solution to the GAIP instance.

\begin{proof}
We begin with the case \(\ID = S\). We note that in this case, \(\D\) will only return \(x \neq \bot\) as a result of lines \ref{ln:fig_PSIReductionS_if_1} or \ref{ln:fig_PSIReductionS_if_2} of Fig.~\ref{fig:PSIReductionS}. Suppose that \(x \neq \bot\) is returned. Then, since the assertion on line \ref{ln:fig_PSIReductionS_assert_1} has passed, we know that there is a unique \(Y \in \mathcal{E}\ell\ell_{\mathbb{F}_{p}}(\mathfrak{O})\) such that
\[
    L[Y \,\|\, m^*] = h^* = L[ [g^{z^*+\sk_V}] \ast \pk_{S,b} \,\|\, m^*]
\]
(where the second equality is because \(h^*\) is defined by the call to \(\prot.\Simul\) on line \ref{ln:fig_PSIReductionS_simul}). Since entries in \(L\) are defined only by (direct or indirect) calls to \(\H\), this means that \(Y = [g^{z^*+\sk_V}] \ast \pk_{S,b}\). Now, because the assertion on line \ref{ln:fig_PSIReductionS_assert_2} has passed, we know that \(\A\) has provided \(Y\) as an output or oracle query, and so we have \(A \in \{E, \pk_{S,0}, \pk_{S,1}, \pk_V\}\) and 
\begin{equation}
\label{eqn:PSI_S_AGAM}
[\g^a] \ast A = Y = [\g^{z^*+\sk_V}] \ast \pk_{S,b}.
\end{equation}
Recall that \(\pk_{S,b} = [\g^{b\cdot r}] \ast X\). We now consider two cases:
\begin{enumerate}
\item If \(A = E\), then~\Cref{eqn:PSI_S_AGAM} implies
\[
[\g^a] \ast E = [\g^{z^* + \sk_V}] \ast \pk_{S,b} \implies [\g^{a-z^*-\sk_V - b\cdot r}] \ast E = X,
\]
so that \(\D\)'s output \(x = a-z^*-\sk_V - b\cdot r\) is a solution to the GAIP instance \(X\).
\item If \(A = \pk_V\), then~\Cref{eqn:PSI_S_AGAM} implies
\[
[\g^a] \ast \pk_V = [\g^{z^* + \sk_V}] \ast \pk_{S,b} \implies [\g^{a-z^*-b\cdot r}] \ast E = X,
\]
so that \(\D\)'s output \(x = a-z^*-b\cdot r\) is a solution to the GAIP instance \(X\).
\end{enumerate}
When \(\ID = V\) the argument is similar. \(\D\) will only return \(x \neq \bot\) as a result of lines \ref{ln:fig_PSIReductionV_if_1}, \ref{ln:fig_PSIReductionV_if_2}, or \ref{ln:fig_PSIReductionV_if_3} of Fig.~\ref{fig:PSIReductionV}. Suppose that \(x \neq \bot\) is returned. Then, since the assertion on line \ref{ln:fig_PSIReductionV_assert_1} has passed, we know that there is a unique \(Y \in \mathcal{E}\ell\ell_{\mathbb{F}_{p}}(\mathfrak{O})\) such that
\[
    L[Y \,\|\, m^*] = h^* = L[ [g^{z^*+\sk_{S,b}}] \ast \pk_{V} \,\|\, m^*]
\]
(where the second equality is because \(h^*\) is defined by the call to \(\prot.\Sign\) on line \ref{ln:fig_PSIReductionV_sign}). Since entries in \(L\) are defined only by (direct or indirect) calls to \(\H\), this means that \(Y = [g^{z^*+\sk_{S,b}}] \ast \pk_{S,b}\). Now, because the assertion on line \ref{ln:fig_PSIReductionV_assert_2} has passed, we know that \(\A\) has provided \(Y\) as an output or oracle query, and so we have \(A \in \{E, \pk_{S,0}, \pk_{S,1}, \pk_V\}\) and 
\begin{equation}
\label{eqn:PSI_V_AGAM}
[\g^a] \ast A = Y = [\g^{z^*+\sk_{S,b}}] \ast \pk_V.
\end{equation}
Recall that \(\pk_V = X\). We now consider three cases:
\begin{enumerate}
\item If \(A = E\), then~\Cref{eqn:PSI_V_AGAM} implies
\[
[\g^a] \ast E = [\g^{z^* + \sk_{S,b}}] \ast \pk_{V} \implies [\g^{a-z^*-\sk_{S,b}}] \ast E = X,
\]
so that \(\D\)'s output \(x = a-z^*-\sk_{S,b}\) is a solution to the GAIP instance \(X\).
\item If \(A = \pk_{S,b}\), then~\Cref{eqn:PSI_V_AGAM} implies
\[
[\g^a] \ast \pk_{S,b} = [\g^{z^* + \sk_{S,b}}] \ast \pk_{V} \implies [\g^{a-z^*}] \ast E = X,
\]
so that \(\D\)'s output \(x = a-z^*\) is a solution to the GAIP instance \(X\).
\item If \(A = \pk_{S,1-b}\), then~\Cref{eqn:PSI_V_AGAM} implies
\[
[\g^a] \ast \pk_{S,1-b} = [\g^{z^* + \sk_{S,b}}] \ast \pk_{V} \implies [\g^{a-z^* -\sk_{S,b} + \sk_{S,1-b}}] \ast E = X,
\]
so that \(\D\)'s output \(x = a-z^* -\sk_{S,b} + \sk_{S,1-b}\) is a solution to the GAIP instance \(X\).
\end{enumerate}
This completes the proof of the claim.\qedhere
\end{proof}
\end{claim}

\begin{claim}
\(\D^S\) and \(\D^V\) perfectly simulate the SUF-CMA game to \(\A\). In particular: fix public parameters \( (p, [\g], N, E, \H)\) for \(\prot\). Let \(\mathrm{View}(\A)\) denote the adversary's \emph{view} in the PSI game---more precisely, \(\mathrm{View}(\A)\) is the joint distribution of the public values \( \pk_S, \pk_V\) and all oracle responses given to \(\A\) in the course of the PSI experiment. Similarly, let \(\mathrm{View}_S(\A)\) and \(\mathrm{View}_V(\A)\) denote the adversary's view in the reductions \(\D^S\) and \(\D^V\), respectively, when \(X\) is a uniformly-chosen GAIP instance. Then, the distributions \(\mathrm{View}(\A), \mathrm{View}_S(\A)\), and \(\mathrm{View}_V(\A)\) are identical.
\end{claim}

The proof of this claim is similar to the analogous claim in the proof of SUF-CMA security in the AGAM, and so we omit it. As a consequence of the claim, the adversary's behaviour in the PSI experiment is identical to its behaviour when interacting with \(\D\).

\begin{claim}
Let \(\A\) be an algebraic adversary for the PSI game against \(\prot\). Suppose that \(\A\) runs in time \(T\) and makes at most \(q\) total queries to \(\H, \OSimul, \OVerify\), and \(\OVerify'\). Define
\[
\Succ^{\D^\A}(\lambda) = \PP\left[x' \neq \bot \middle| \begin{array}{rl}
(p, \g, N, E, \H) &\gets \Setup(1^\lambda) \\
x &\sample \ZZ_N \\
X &\gets [\g^x] \ast E \\
x' &\gets \D^\A(X)
\end{array}
\right].
\]
Then there exists a GAIP algorithm \(\D\) running in time \(T + \poly[\lambda]\) such that 
\[
\Succ^{\D^\A}(\lambda) \geq \frac{1}{2}\Adv^{\PSI,\A}_{\prot}(\lambda) - \frac{1}{2}\Adv^{\mathrm{GAIP},\C}(\secpar) - \frac{(q+1)^2 + q_\Verify}{2^{2\lambda+1}} - \frac{q_\Simul + q_\Verify}{2N}.
\]

\begin{proof}
Since \(\ID\) is sampled uniformly at random from \(\{S,V\}\) in the execution of \(\D^\A\), we have
\begin{equation}
\label{eqn:PSI_B_Succ}
\Succ^{\D^\A}(\lambda) = \frac{1}{2}\PP[\D^\A(X) \neq \bot | \ID = S] + \frac{1}{2}\PP[\D^\A(X) \neq \bot | \ID = V]
\end{equation}
and so it suffices to analyze the success probability of \(\D^S\) and \(\D^V\) individually. Beginning with \(\D^S\) and referring to Fig.~\ref{fig:PSIReductionS}, we see that \(\D^S\) succeeds as long as:
\begin{enumerate}
\item The assertion on line \ref{ln:fig_PSIReductionS_assert_1} passes.
\item The assertion on line \ref{ln:fig_PSIReductionS_assert_2} passes.
\item When \( (a,A)\) is constructed on line \ref{ln:fig_PSIReductionS_GetRep}, we have \(A \not\in \{\pk_{S,0}, \pk_{S,1}\}\). 
\end{enumerate}
For the first condition, first observe that some such \(Y\) must exist, because it is defined by the call to \(\prot.\Simul\) on line \ref{ln:fig_PSIReductionS_simul}  (if not earlier). By the birthday bound, the probability that two or more such values of \(Y\) satisfy the condition is at most \( \frac{(q+1)^2}{2^{2\lambda}}\), since at most \( q+1\) hash values are defined (\(q\) because of \(\A\)'s queries to the various oracles, and 1 by the challenge signature construction on line \ref{ln:fig_PSIReductionS_simul}).

For the second condition, Lemma~\ref{lem:AGAM_PSI_Queries} guarantees that \(\A\) makes a query to \(\H\) at \( (Y \,\|\, m^*)\) with \(Y = [\g^{z^*+\sk_V}] \ast \pk_{S,b}\) with probability at least
\begin{equation}
\label{eqn:PSI_Query_Bound}
\Adv^{\PSI,\A}_{\prot}(1^\lambda) - \Adv^{\mathrm{GAIP}, \C}(1^\lambda) - \frac{q_\Simul + q_\Verify}{N} - \frac{q_\Verify}{2^{2\lambda}}
\end{equation}
where \(\C\) is a GAIP adversary running in time \(T + \poly[\lambda]\). Given that the assertion on line \ref{ln:fig_PSIReductionS_assert_1} passes, \(\A\) making such a query will ensure that the assertion on line \ref{ln:fig_PSIReductionS_assert_2} passes, since the assertion on line \ref{ln:fig_PSIReductionS_assert_1}, combined with the definition of the challenge signature on line \ref{ln:fig_PSIReductionS_simul} guarantees that the only \(Y\) such that \(h^* = L[Y \,\|\, m^*]\) is precisely 
\[
Y = [\g^{z^*+\sk_{S,b}}] \ast \pk_V = [\g^{z^*+\sk_{V}}] \ast \pk_{S,b} 
\]
where the first equality from the definition of \(\prot.\Sign\) and the construction of \(\sigma^*\) on line \ref{ln:fig_PSIReductionS_simul}, and the second comes from the commutativity of the group action. Thus, the probability that both assertions pass is at least
\[
\Adv^{\PSI,\A}_{\prot}(1^\lambda)- \Adv^{\mathrm{GAIP}, \C}(1^\lambda) - \frac{q_\Simul + q_\Verify}{N} - \frac{q_\Verify}{2^{2\lambda}} - \frac{(q+1)^2}{2^{2\secpar}}.
\]

When both assertions pass, we know that \(A \in \{E, \pk_{S,0}, \pk_{S,1}, \pk_V\}\). Let \(p_{E}, p_{S,0}, p_{S,1}\) and, \(p_V\) denote the probability that, when both assertions pass, that \(A = E, \pk_{S,0}, \pk_{S,1}\), and \(\pk_V\), respectively. Then
\begin{align*}
\PP[\mbox{Assertions pass}] &= p_E + p_{S,0} + p_{S,1} + p_V\\
& \geq \Adv^{\PSI,\A}_{\prot}(1^\lambda)- \Adv^{\mathrm{GAIP}, \C}(1^\lambda) - \frac{q_\Simul + q_\Verify}{N} - \frac{(q+1)^2+q_\Verify}{2^{2\secpar}}.
\end{align*}
When computing \(\PP[\D^\A(1^\lambda) \neq \bot | \ID = S]\), we must subtract \(p_{S,0} + p_{S,1}\), since the corresponding values of \(A\) do not allow \(\D^S\) to succeed. What remains is
\[
\PP[\D^\A(1^\lambda) \neq \bot | \ID = S] = p_E + p_V.
\]
A similar argument yields
\[
\PP[\D^\A(1^\lambda) \neq \bot | \ID = S] = p_E + p_{S,0} + p_{S,1}
\]
where, as in the AGAM proof of SUF-CMA security, the probabilities \(p_E, p_V, p_{S,0}, p_{S,1}\) do not change between \(\D^S\) and \(\D^V\). By Equations~(\ref{eqn:PSI_B_Succ}) and~(\ref{eqn:PSI_Query_Bound}), we have
\begin{align}
\nonumber
\Succ^{\D^\A}(\lambda) &= p_E + \frac{1}{2} (p_{S,0} + p_{S,1} + p_V)\\ 
&\geq \frac{1}{2}\Adv^{\PSI,\A}_{\prot}(\lambda) - \frac{1}{2}\Adv^{\mathrm{GAIP},\C}(\secpar) - \frac{(q+1)^2 + q_\Verify}{2^{2\lambda+1}} - \frac{q_\Simul + q_\Verify}{2N} 
\label{eqn:AGAM_PSI_Bound}
\end{align}
as required. \qedhere
\end{proof}
\end{claim}

Finally, since $\Succ^{\D^\A}(\lambda) = \Adv^{\mathrm{GAIP},\D}(\secpar)$. Equation (6) implies that 
\[
\Adv^{\mathrm{GAIP},\D}(\secpar) + \frac{1}{2}\Adv^{\mathrm{GAIP},\C}(\secpar) \geq \frac{1}{2}\Adv^{\PSI,\A}_{\prot}(\lambda) - \frac{(q+1)^2 + q_\Verify}{2^{2\lambda+1}} - \frac{q_\Simul + q_\Verify}{2N} 
\]
Taking the maximum of $\Adv^{\mathrm{GAIP},\D}(\secpar)$ and $\Adv^{\mathrm{GAIP},\C}(\secpar) $, and calling the corresponding algorithm $\E$, we have 
\begin{align*}
    \frac{3}{2} \Adv^{\mathrm{GAIP},\E}(\secpar) & \geq  \frac{1}{2}\Adv^{\PSI,\A}_{\prot}(\lambda) - \frac{(q+1)^2 + q_\Verify}{2^{2\lambda+1}} - \frac{q_\Simul + q_\Verify}{2N} \\[6pt]
   \implies \Adv^{\mathrm{GAIP},\E}(\lambda) & \geq \frac{1}{3}\Adv^{\PSI,\A}_{\prot}(\lambda) - \frac{(q+1)^2 + q_\Verify}{3\cdot 2^{2\lambda}} - \frac{q_\Simul + q_\Verify}{3N}.
\end{align*}

\begin{proof}[Proof (of Theorem~\ref{thm:AGAM_PSI})]

Under the GAIP assumption, if \(\A\) is polynomial-time, \(\Adv^{\mathrm{GAIP},\E}(\secpar)\), and \(\frac{(q+1)^2 + q_\Verify}{3\cdot 2^{2\lambda}} + \frac{q_\Simul + q_\Verify}{3N}\) are negligible. Therefore, Lemma~\ref{lem:QueryImpliesGAIP} forces that \( \Adv^{\PSI,\A}_{\prot}(\lambda) \) to be negligible.
\end{proof}

\section{Performance}\label{sec:efficiency}

Table~\ref{tab:comparison} summarizes the size complexity of our protocol when instantiated with the CSIDH-512 parameter set.

\begin{description}\itemsep3pt

\item[{Secret Keys:}]
The secret keys $\sk_S$ and $\sk_V$ are sampled uniformly from $\mathbb{Z}_N$, where $\log_2 N = 2\lambda$. Hence, each secret key requires $2\lambda$ bits.

\item[{Public Keys:}]
Each elliptic curve in $\mathcal{E}\ell\ell_{\mathbb{F}_p}(\mathfrak{O})$ is represented in Montgomery form $E_A: y^2 = x^3 + Ax^2 + x$, where $A \in \mathbb{F}_p$. Since $p$ is a $4\lambda$-bit prime, each public key consists of a single field element in $\mathbb{F}_p$, which requires $4\lambda$ bits.

\item[{Signatures:}]
A signature $\sigma = (h, z)$ consists of a $2\lambda$-bit hash value and an element $z \in \mathbb{Z}_N$, where the hash output length is chosen to match the order of $\mathbb{Z}_N$. Hence, the total signature size is $4\lambda$ bits.

\end{description}

\begin{table}[h!]
\vspace{-0.5cm}
\caption{Key and signature sizes (in bits) using the CSIDH-512 parameter set.}
\label{tab:comparison}
\centering
\scriptsize
\setlength{\tabcolsep}{2pt}

\begin{tabular*}{\linewidth}{@{\extracolsep{\fill}}lcccc}
\toprule
\textbf{Scheme} & \textbf{Secret Key} & \textbf{Public Key} &
\textbf{Signature} & \textbf{Security} \\
\midrule

Sun et al.~\cite{SIDH_SDVS}\({}^*\)
& 432 & 2\,640 & 256
& SS-DDH, ROM \\

Khuc et al.~\cite{khuc2025}
& 128 & 131\,072 & 6\,896
& CSI-CDH, ROM \\

$\mathbf{\mathsf{CSI\text{-}SDVS}}$
& \textbf{256} & \textbf{512} & \textbf{512}
& \textbf{CSI-DDH, ROM} \\
& & & & or \textbf{GAIP, AGAM, ROM}\\
\bottomrule
\end{tabular*}

\vspace{0.5em}

\parbox{\linewidth}{\footnotesize
\textit{Note.}
\({}^*\)This scheme is insecure;
ROM: Random Oracle Model;
GAIP: Group Action Inverse Problem;
CSI-CDH: Commutative Supersingular Isogeny-Computational Diffie--Hellman Problem;
SS-CDH: Supersingular Computational Diffie--Hellman Problem;
SS-DDH: Supersingular Decisional Diffie--Hellman Problem.
}
\vspace{-0.5cm}
\end{table}

While prior schemes exhibit large public keys or signatures, $\mathsf{CSI\text{-}SDVS}$ achieves balanced $\mathcal{O}(\lambda)$-size representations across all components.



\begin{thebibliography}{100}

\bibitem{DVS1}
Jakobsson, M., Sako, K., Impagliazzo, R.: Designated verifier proofs and their applications.
In: International Conference on the Theory and Applications of Cryptographic Techniques,
pp. 143--154 (1996)

\bibitem{DVS2}
Chaum, D.: Private signature and proof systems.
Google Patents, US Patent 5,493,614 (1996)

\bibitem{SDVS}
Laguillaumie, F., Vergnaud, D.: Designated verifier signatures: Anonymity and efficient
construction from any bilinear map.
In: International Conference on Security in Communication Networks,
pp. 105--119 (2004)

\bibitem{SDVS2}
Saeednia, S., Kremer, S., Markowitch, O.: An efficient strong designated verifier signature scheme.
In: International Conference on Information Security and Cryptology,
pp. 40--54 (2003)

\bibitem{shortSDVS}
Huang, X.-Y., Susilo, W., Mu, Y., Zhang, F.-T.: Short designated verifier signature scheme
and its identity-based variant.
International Journal of Network Security 6(1), 82--93 (2008)

\bibitem{idSDVS}
Kang, B., Boyd, C., Dawson, E.D.: A novel identity-based strong designated verifier signature scheme.
Journal of Systems and Software 82(2), 270--273 (2009)

\bibitem{Shor}
Shor, P.W.: Polynomial-time algorithms for prime factorization and discrete logarithms
on a quantum computer. SIAM Review 41(2), 303--332 (1999)

\bibitem{assar2022}
Asaar, M.R.: Code-based strong designated verifier signatures: Security analysis and a new construction.
Cryptology ePrint Archive (2016)

\bibitem{ren2022}
Ren, Y., Wang, H., Du, J., Ma, L.: Code-based authentication with designated verifier.
International Journal of Grid and Utility Computing 7(1), 61--67 (2016)

\bibitem{shooshtari2021}
Shooshtari, M.K., Ahmadian-Attari, M., Aref, M.R.: Provably secure strong designated verifier
signature scheme based on coding theory.
International Journal of Communication Systems 30(7), e3162 (2017)

\bibitem{Thanalakshmi}
Thanalakshmi, P., Anitha, R.: A new code-based designated verifier signature scheme.
International Journal of Communication Systems 31(17), e3803 (2018)

\bibitem{khuc2025}
Khuc, X.T., Susilo, W., Duong, D.H., Li, Y., Roy, P.S., Fukushima, K., Kiyomoto, S.:
Strong designated verifier signatures from isogeny assumptions.
In: International Conference on Cryptology and Network Security,
pp. 46--69 (2025)

\bibitem{AGAM}
Duman, J., Hartmann, D., Kiltz, E., Kunzweiler, S., Lehmann, J., Riepel, D.:
Generic models for group actions.
In: IACR International Conference on Public-Key Cryptography,
pp. 406--435 (2023)

\bibitem{AGM}
Fuchsbauer, G., Kiltz, E., Loss, J.: The algebraic group model and its applications.
In: Annual International Cryptology Conference,
pp. 33--62 (2018)

\bibitem{sidh}
Jao, D., De Feo, L.: Towards quantum-resistant cryptosystems from supersingular elliptic curve isogenies.
In: International Workshop on Post-Quantum Cryptography,
pp. 19--34 (2011)

\bibitem{attack1SIDH}
Castryck, W., Decru, T.: An efficient key recovery attack on SIDH.
In: Annual International Conference on the Theory and Applications of Cryptographic Techniques,
pp. 423--447 (2023)

\bibitem{attack2SIDH}
Maino, L., Martindale, C., Panny, L., Pope, G., Wesolowski, B.:
A direct key recovery attack on SIDH.
In: Annual International Conference on the Theory and Applications of Cryptographic Techniques,
pp. 448--471 (2023)

\bibitem{attack3SIDH}
Robert, D.: Breaking SIDH in polynomial time.
In: Annual International Conference on the Theory and Applications of Cryptographic Techniques,
pp. 472--503 (2023)

\bibitem{csidh}
Castryck, W., Lange, T., Martindale, C., Panny, L., Renes, J.:
CSIDH: An efficient post-quantum commutative group action.
In: Advances in Cryptology--ASIACRYPT 2018: 24th International Conference on the
Theory and Application of Cryptology and Information Security,
pp. 395--427 (2018)

\bibitem{PEGASIS}
Dartois, P., Eriksen, J.K., Fouotsa, T.B., Herl{\'e}dan Le Merdy, A.,
Invernizzi, R., Robert, D., Rueger, R., Vercauteren, F., Wesolowski, B.:
PEGASIS: Practical effective class group action using 4-dimensional isogenies.
In: Annual International Cryptology Conference,
pp. 67--99 (2025)

\bibitem{GPV}
Gentry, C., Peikert, C., Vaikuntanathan, V.:
Trapdoors for hard lattices and new cryptographic constructions.
In: Proceedings of the Fortieth Annual ACM Symposium on Theory of Computing,
pp. 197--206 (2008)

\bibitem{lattice1}
Wang, F., Hu, Y., Wang, B.: Lattice-based strong designate verifier signature and its applications.
Malaysian Journal of Computer Science 25(1), 11--22 (2012)

\bibitem{lattice2}
Noh, G., Jeong, I.R.: Strong designated verifier signature scheme from lattices in the standard model.
Security and Communication Networks 9(18), 6202--6214 (2016)

\bibitem{lattice3}
Cai, J., Jiang, H., Zhang, P., Zheng, Z., Lyu, G., Xu, Q.:
An efficient strong designated verifier signature based on $\mathcal{R}$-SIS assumption.
IEEE Access 7, 3938--3947 (2019)

\bibitem{lattice4}
Zhang, Y., Susilo, W., Guo, F.: Lattice-based strong designated verifier signature
with non-delegatability.
Computer Standards \& Interfaces 92, 103904 (2025)

\bibitem{SIDH_SDVS}
Sun, X., Tian, H., Wang, Y.: Toward quantum-resistant strong designated verifier signature from isogenies.
In: 2012 Fourth International Conference on Intelligent Networking and Collaborative Systems,
pp. 292--296 (2012)

\bibitem{washington}
Washington, L.C.: Elliptic Curves: Number Theory and Cryptography.
Chapman and Hall/CRC, Boca Raton, FL (2008)

\bibitem{silverman}
Silverman, J.H.: The Arithmetic of Elliptic Curves.
Springer, New York, NY, vol. 106 (2009)


\bibitem{alamati_CGA}
Alamati, N., De Feo, L., Montgomery, H., Patranabis, S.:
Cryptographic group actions and applications.
In: International Conference on the Theory and Application of Cryptology and Information Security,
pp. 411--439 (2020)

\end{thebibliography}

\end{document}